\documentclass[11pt,a4paper]{article}
\pdfoutput=1
\usepackage{jcappub}
\usepackage{graphicx}
\usepackage{dcolumn}
\usepackage{bm}
\usepackage{latexsym}
\usepackage{amsfonts}
\usepackage{amssymb}
\usepackage{amsmath}
\usepackage{mathrsfs}
\usepackage{afterpage}
\usepackage{placeins}
\usepackage{color}
\usepackage{hyperref}
\usepackage{ulem}
\usepackage{natbib} 
\usepackage{verbatim}
\usepackage[left=2.5cm,right=2cm,top=3cm,bottom=2.5cm]{geometry}

\begin{document}

\title{Parity breaking signatures from a Chern-Simons coupling during inflation: the case of non-Gaussian gravitational waves
}

\author[a,b,c]{Nicola Bartolo,}
\author[a,b]{Giorgio Orlando}
\emailAdd{nicola.bartolo@pd.infn.it, giorgio.orlando@phd.unipd.it}

\affiliation[a]{
 Dipartimento di Fisica e Astronomia ``G. Galilei", Universit\`{a} degli Studi di Padova, via Marzolo 8, I-35131, Padova, Italy}
\affiliation[b]{
 INFN Sezione di Padova, via Marzolo 8, I-35131, Padova, Italy
 }
\affiliation[c]{
 INAF-Osservatorio Astronomico di Padova, Vicolo dell'Osservatorio 5, I-35122 Padova, Italy}


\abstract{Considering high-energy modifications of Einstein gravity during inflation is an interesting issue. We can constrain the strength of the new gravitational terms through observations of inflationary imprints in the actual universe. In this paper we analyze the effects on slow-roll models due to a Chern-Simons term coupled to the inflaton field through a generic coupling function $f(\phi)$. A well known result is the polarization of primordial gravitational waves (PGW) into left and right eigenstates, as a consequence of parity breaking. In such a scenario the modifications to the power spectrum of PGW are suppressed under the conditions that allow to avoid the production of ghost gravitons at a certain energy scale, the so-called Chern-Simons mass $M_{CS}$.
 In general it has been recently pointed out that there is very little hope to efficiently constrain chirality of PGW on the basis solely of two-point statistics from future CMB data, even in the most optimistic cases. Thus we search if significant parity breaking signatures can arise at least in the bispectrum statistics. We find that the tensor-tensor-scalar bispectra $\langle \gamma \gamma \zeta \rangle$ for each polarization state are the only ones that are not suppressed. Their amplitude, setting the level of parity breaking during inflation, is proportional to the second derivative of the coupling function $f(\phi)$ and they turn out to be maximum in the squeezed limit. We comment on the squeezed-limit consistency relation arising in the case of chiral gravitational waves, and on possible observables to constrain these signatures.}

\maketitle

\section{Introduction}

\noindent Slow-roll models of inflation, the actual inflationary paradigm~\cite{Brout:1977,Kazanas:1980, Starobinsky:1980, Mukhanov:1981,Guth:1980,Sato:1980,Linde:1981}, describe successfully many observed features of the universe, its homogeneity, flatness, and in particular the origin of the first  curvature (density) perturbations~\citep{Planck_results_2016, Planck_constriant_inflation_2016,Planck_primordial_NG_2016}. In most of such models Einstein gravity is usually assumed to describe the theory of gravity. However, because of the possibly very high energies underlying the inflationary dynamics, it might be that remnant signatures of modification to Einstein gravity are left imprinted in the inflationary quantum fluctuations.
 The examples are different, from the first model of inflation~\cite{Starobinsky:1980}, based on $R^2$-higher order gravitational terms, to more recent scenarios, such as, e.g.,~\cite{Weinberg:2008, Guo:2010,Kobayashi:2010cm,Burrage:2010cu,Kobayashi:2011nu,DeFelice:2011,Deruelle:2011,Guo:2013,Bartolo:2014,Xia:2015, Myrzakul:2015, Kanti:2015, Arkani:2015,Baumann:2016, Lee:2016, Nozari:2016, Fomin:2017, Nozari:2017, Alexander:2005, Satoh:2008,Alexander:2009, Satoh:2010, Myung:2014, Alexander:2016, Yong:2016,Maldacena:2011, Soda:2011, Shiraishi:2011,Zhu:2013fja}.~\footnote{For more general inflationary frameworks, see also, e.g.,~\cite{Noumi:2014zqa,Creminelli:2014wna,Endlich:2012pz,Cannone:2014uqa,Bartolo:2015qvr,Ricciardone:2016lym,Bordin:2017, Dimastrogiovanni:2014, Dimastrogiovanni:2016}. For a review on inflationary gravitational waves containting other examples, see~\cite{Guzzetti:2016mkm}.} In such modified gravity models of inflation we can have modifications in both the background dynamics and the evolution of primordial perturbations. In particular production of primordial non-Gaussianities is an interesting feature of these models. In fact, in the simplest models of inflation with standard gravity, the non-linearity parameter $f_{NL}$, which measures the level of non-Gaussianity in the primordial curvature perturbations, is of the order $\sim \mathcal{O}(\epsilon, \eta)$~\cite{Acquaviva:2003, Maldacena:2003}; for this reason in the standard scenario primordial non-Gaussianity is highly suppressed by slow-roll parameters. Instead, in a modified gravity model we can have enhancement of non-Gaussianities (see e.g.~\cite{Bartolo:2014} and Refs. therein). This is due to the fact that a modification of Einstein gravity can bring new degrees of freedom as well as new interactions among fundamental fields of the theory. 
 
In this paper we will analyse modifications in slow-roll models of inflation provided by Chern-Simons gravity. In the literature effects of this kind of gravity on gravitational waves in an inflationary context have been studied for the first time in Ref.~\cite{Lue:1999} , and in Ref.~\cite{Jackiw:2003} more details have been elaborated ​on more general aspects of the theory, while more recent works include Refs.~\cite{Alexander:2005,Alexander:2006, Alexander:2007, Satoh:2008, Satoh:2010,Maleknejad:2012, Myung:2014, Alexander:2016, Yong:2016, Kawai:2017}. This theory consists in a parity-breaking modification of standard Einstein gravity in which one adds the so-called Chern-Simons gravitational term coupled to the inflaton field in the Lagrangian of the slow-roll inflationary models. This term can be written as 
\begin{equation}\label{Chern-Simons}
\Delta \mathcal L = f(\phi)  \epsilon^{\mu\nu\rho\sigma} {R_{\mu \nu}}^{\kappa \lambda}R_{\rho \sigma\kappa\lambda} \mbox{ ,}
\end{equation}
where $f(\phi)$ is a generic function of the scalar inflaton field $\phi$ only, $\epsilon^{\mu\nu\rho\sigma}$ is the total antisymmetric Levi-Civita pseudo-tensor (with $\epsilon^{1230}= 1$), $R_{\mu \nu \rho \sigma}$ is the Riemann curvature tensor. In Eq. \eqref{Chern-Simons} one can replace the Riemann tensor with the Weyl tensor, Eq.~\eqref{Weyl_tensor}, without introducing any modification to the Lagrangian (see, e.g. \cite{Grumiller:2007rv}, and Appendix \ref{appendix_A}). For this reason the Chern-Simons gravitational term belongs to the so-called Weyl-square terms  and it is abbreviated with the symbol $\tilde C C$. The form of the coupling function $f(\phi)$ could be determined a priori by a quantum theory of gravity, which is still incomplete for the moment. For this reason, we can leave the coupling totally general.  A fundamental result is the polarization of primordial gravitational waves (PGW) into chiral-eigenstates, the so-called left (L) and right (R) polarization states and the possibility to detect such parity breaking signatures looking at CMB polarization~\cite{Lue:1999}. The reason is that a parity transformation changes the chirality of PGW, and so in a theory in which parity is broken we expect a difference in the dynamical evolution of these chiral-eigenstates. In addition, the term \eqref{Chern-Simons} introduces in the theory a new mass scale, the so-called Chern-Simons mass $M_{CS}$. At energies greater than $M_{CS}$ we have formation of ghost gravitons in the physical spectrum (see e.g. Ref. \cite{Dyda:2012}). The parity breaking in PGW power spectra is a well known result (see e.g. Refs \cite{Lue:1999,Satoh:2010, Alexander:2005, Alexander:2016}). In particular in Ref.~\cite{Satoh:2010} parity breaking signatures are studied in a regime of energies where one can neglect the presence of ghost fields: in this case parity breaking signature in the power-spectra is proportional to the first derivative of the coupling $f(\phi)$ w.r.t. the inflaton, namely $f'(\phi)$. However, such a signature is predicted to be small due to the theoretical assumption (i.e. the absence of ghost fields) which constrains the value of $f'(\phi)$. 

There are several observational studies on parity breaking signatures of PGW (and on cosmological birefringence) using CMB data~\cite{Maravin:2008,Feng:2006,Xia:2008,Ichiki:2007,Cabella:2007,Gubitosi:2009,Das:2009,Gruppuso:2012,Gluscevic:2012,Gubitosi:2013,Kaufman:2014,Kahniashvili:2014,Gubitosi:2014,Galaverni:2015,Planck_coll_2016,Polarbear_coll_2015,Gruppuso:2016,Planck_parity_breaking_2016}. At present current CMB power spectra are not able to efficiently constrain the level of parity breaking. Moreover it has been shown recently, on very general ground, that even from future CMB data it will be almost hopeless to constrain the chirality of PGW exploiting only the two-point correlation statistics, specifically the cross-correlation of B-modes with temperature T anisotropies and E-polarization mode~\cite{Gerbino:2016}. Only in the most optimistic cases some weak constraints might be placed. Therefore it has become even more interesting and crucial to ask whether relevant parity breaking signatures can arise in higher-order correlators.

In this paper we will investigate parity breaking signatures in primordial bispectra that comes from the term \eqref{Chern-Simons}. As in Ref. \cite{Satoh:2010}, we will work within an energy range in which ghost fields are absent.  We will analyze the tensor fluctuations (gravitational waves) bispectra $\langle \gamma \gamma \gamma \rangle$ and their mixed correlators with the scalar curvature perturbation $\zeta$, $\langle \gamma \zeta \zeta \rangle$ and $\langle \gamma \gamma \zeta \rangle$. Analyzing the scalar bispectrum $\langle \zeta \zeta \zeta \rangle$ is not interesting for our goals, because it is insensitive to parity breaking signatures, being sourced only by scalar fields. In particular we will focus on the bispectrum $\langle \gamma \gamma \zeta \rangle$. In fact, we will show that contributions to other bispectra are proportional to $f'(\phi)$ and so they are suppressed as the power-spectrum case. Instead, we will find that the parity breaking signature in the bispectrum $\langle \gamma \gamma \zeta \rangle$ is proportional to the second derivative of the coupling $f(\phi)$. We will show that also in this case the parameter $f''(\phi)$ is constrained by some theoretical assumptions. However, such a constraint is less stringent than the power-spectrum case. The result is that in our model a possibly relevant parity breaking signature in the bispectrum $\langle \gamma \gamma \zeta \rangle$ is compatible with slow-roll theories of inflation.

In Refs.~\cite{Maldacena:2011, Soda:2011, Shiraishi:2011} possible parity violating signatures in graviton non-Gaussianities due to a Weyl-cubic parity breaking term ($\tilde C C^2$) have been considered. As we show in details there are various differences w.r.t. the case we investigate in this paper. For example, notice that for a Weyl-cubic term $\tilde C C^2$ no coupling function with the inflaton is required to have a non trivial contribution, since it is enough to consider corrections in slow-roll parameter $\epsilon$. This is particularly different by our case because a Weyl-square term ($\tilde C C$) can be the source of a non trivial parity breaking signature in the mixed correlator $\langle \gamma \gamma \zeta \rangle$ only if it is coupled to the inflaton. Otherwise it would provide a surface term and it can be dropped by the total Lagrangian.

The paper is organized as follows. In Section \ref{2} we recall how the Chern-Simons term  might originate from a generalization of Einstein theory of gravity. Moreover we describe some peculiar features of this term and we write down the action of the theory. In Section \ref{3} we describe the theoretical formalism which we have adopted in performing the computations, specifying the form of the metric tensor and the gauge fixing adopted. In Section \ref{4} we review briefly the results about parity breaking signatures in the power-spectra of tensor perturbations in the case in which we assume no ghost field production during inflation. Under the same conditions, in Section \ref{5}, we analyze possible parity breaking signatures in primordial bispectra involving tensor modes. In particular a detailed computation regarding the bispectrum $\langle \gamma \gamma \zeta \rangle$ is showed. In Section \ref{6}, we make some comments about the so-called consistency relations in the squeezed limit of this inflationary bispectrum. Finally in Section \ref{7} we draw our conclusions and mention some possible observational ways of measuring the parity breaking signatures discussed in this paper.

\section{Origin of the Chern-Simons term}\label{2}

\noindent The Hilbert-Einstein (H-E) action is built by admitting covariant terms with a maximum number of two derivatives of the metric tensor. The only covariant term which obeys to this constraint is the scalar curvature $R$ and the result is the density Lagrangian
\begin{equation}\label{HE}
\mathcal{L}_{HE}= \sqrt{g} \left(\frac{M^2_{Pl}}{2} R \right) , 
\end{equation}
where $g= -det [g_{\mu \nu}]$ and $M^2_{Pl}= (8 \pi G)^{-1}$ is the reduced Planck mass.  Here $\sqrt{g}$ is part of the covariant integration measure in the action.

A standard way to modify Einstein gravity is to relax this condition and consider a more general theory of gravity in which the action is built with an expansion in series of covariant terms that contains an increasing number of derivatives of the metric tensor. We follow here the approach of Ref.~\cite{Weinberg:2008}. The first correction to the Lagrangian \eqref{HE} is obtained by considering the most general covariant terms with four derivatives of the metric tensor. We build these terms doing tensor contractions between two tensors with two derivatives of the metric tensor. Such tensors are the fundamental curvature tensors of General Relativity: the Riemann tensor $R_{\mu \nu \rho \sigma}$, the Ricci tensor $R_{\mu \nu}$ and the scalar curvature $R$. 
Then, the most general expression for the additional Lagrangian we want to focus on is
\begin{equation}\label{effective_lagrangian}
\Delta \mathcal{L}= \sqrt{g} \left(f_1 R^2 + f_2 R_{\mu \nu} R^{\mu \nu} + f_3 R_{\mu \nu \rho \sigma} R^{\mu \nu \rho \sigma} \right) + f_4 \epsilon^{\mu \nu \rho \sigma} {R_{\mu \nu}}^{\kappa \lambda} R_{\rho \sigma \kappa \lambda} , 
\end{equation}
where $f_n$ are some dimensionless coefficients. From Lagrangian \eqref{effective_lagrangian} we see that the Chern-Simons term (the one which multiplies $f_4$) is appearing naturally by this (effective field theory)  expansion.

It is convenient to rewrite the Lagrangian \eqref{effective_lagrangian} in terms of the Weyl tensor $C_{\mu \nu \rho \sigma}$ which is the traceless part of the Riemann tensor. This because we can understand better some properties of the Chern-Simons term that will be useful to perform the computations later on. The Weyl tensor is defined by the tensor relation
\begin{equation}\label{Weyl_tensor}
C_{\mu \nu \rho \sigma} = R_{\mu \nu \rho \sigma}- \frac{1}{2} \left(g_{\mu \rho} R_{\nu \sigma} - g_{\mu \sigma} R_{\nu \rho} - g_{\nu \rho} R_{\mu \sigma} + g_{\nu \sigma} R_{\mu \rho} \right)+ \frac{R}{6} \left(g_{\mu \rho} g_{\nu \sigma} - g_{\nu \rho} g_{\mu \sigma} \right) .
\end{equation}
\noindent Thus, by doing a simple redefinition of the coefficients $f_n$ (see Appendix \ref{appendix_A}), Lagrangian \eqref{effective_lagrangian} becomes:
\begin{equation}\label{effective_lagrangian_weyl}
\Delta \mathcal{L}= \sqrt{g} \left(f_1 R^2 + f_2 R_{\mu \nu} R^{\mu \nu} + f_3 C_{\mu \nu \rho \sigma} C^{\mu \nu \rho \sigma} \right) + f_4 \epsilon^{\mu \nu \rho \sigma} {C_{\mu \nu}}^{\kappa \lambda} C_{\rho \sigma \kappa \lambda} .
\end{equation}
This if we consider only the metric tensor $g_{\mu \nu}$. If we also introduce a scalar field $\phi$ which plays the role of the inflaton field we should include also covariant terms up to four derivatives of the inflaton field itself. The result is the full Lagrangian (see e.g. Ref. \cite{Weinberg:2008}):
\begin{align} \label{L_general}
\mathcal{L}&=\sqrt{g}\Bigg[\frac{1}{2}M_{Pl}^2 R-\frac{1}{2}g^{\mu\nu}\partial_\mu\phi\partial^\mu\phi-V(\phi)\nonumber +\\
&+f_3(\phi)\Big(g^{\mu\nu}\partial_\mu\phi\partial_\nu\phi\Big)^2+f_4(\phi)g^{\rho\sigma}\partial_\rho\phi\partial_\sigma\phi\Box\phi\nonumber+f_5(\phi)\Big(\Box \phi\Big)^2 +f_7(\phi)R^{\mu\nu}\partial_\mu\phi\partial_\nu\phi \nonumber +\\
&+f_8(\phi)R\, g^{\mu\nu}\partial_\mu\phi\partial_\nu\phi\nonumber +f_9(\phi)R\, \Box \phi +f_{10}(\phi)R^2 +f_{11}(\phi)R^{\mu\nu}R_{\mu\nu}+f_{12}(\phi)C^{\mu\nu\rho\sigma}C_{\mu\nu\rho\sigma}\Bigg]
\nonumber +\\
&+f_{13}(\phi)\epsilon^{\mu\nu\rho\sigma}C_{\mu\nu}{}^{\kappa\lambda}C_{\rho\sigma\kappa\lambda}\, ,
\end{align}
where we allow the various dimensionless coefficients $f_n$ to depend on the inflaton field. In the first line we recognize the standard action of slow-roll models of inflation. In the last line the Chern-Simons term coupled to the inflaton field appears. In this paper we are interested only in signatures of parity breaking modifications of Einstein gravity during the inflationary epoch. For this reason we restrict to the following Lagrangian
\begin{equation} \label{L_studied}
\mathcal{L}=\sqrt{g}\Bigg[\frac{1}{2}M_{Pl}^2 R-\frac{1}{2}g^{\mu\nu}\partial_\mu\phi\partial^\mu\phi-V(\phi) \Bigg]+ f(\phi)\epsilon^{\mu\nu\rho\sigma}C_{\mu\nu}{}^{\kappa\lambda}C_{\rho\sigma\kappa\lambda}\, .
\end{equation}
In any case, in the full Lagrangian \eqref{L_general}, the Chern-Simons term has other interesting peculiarities with respect to the other terms. First of all, the Chern-Simons term is zero if computed on the background metric (the background metric is the Friedmann-Robertson-Walker (FRW) metric, which is conformally invariant and thus the Weyl tensor is zero in FRW). As a consequence, the Chern-Simons term does not modify the background dynamics of inflation. Another feature of the Chern-Simons term is the fact that it is invariant under a Weyl transformation of the metric. This fact will also be useful in performing the computations (also the term $C^{\mu\nu\rho\sigma}C_{\mu\nu\rho\sigma}$ have the same properties, and it has been shown that such a term introduces in the theory ghost fields (see e.g. Ref. \cite{Deruelle:2011})). As a final consideration, we want to emphasize that the Chern-Simons term is a total derivative term (see e.g. Ref. \cite{Jackiw:2004}); for this reason the coupling with the inflaton $f(\phi)$ is necessary to achieve a non trivial signature in the theory. Now, before performing our analysis, we briefly describe the formalism which we have employed.

\section{Formalism adopted}\label{3}

\subsection{Metric}

\noindent We adopt the Arnowitt-Deser-Misner (ADM) formalism of the metric (see e.g. Refs. \cite{Arnowitt:2008, Golovnev:2013}). In Cartesian coordinates metric components are written as
\begin{equation}\label{ADM}
g_{00}= -(N^2- N_i N^i), \quad g_{0i}= N_i, \quad g_{ij}= a^2 h_{ij} \mbox{ ,}
\end{equation}
where $a$ is the scale factor of the universe and $h_{ij}= (1+B)\delta_{ij}+G_{ij}$. Here $B$ and $G_{ij}$ are cosmological perturbations around a FRW background metric and $N$ and $N_i$ are auxiliary fields (in standard gravity).

Here we can decompose the vector $N_i$ and the tensor $G_{ij}$ as:
\begin{align}
N_i= & \partial_i D + E_i \mbox{ ,} \label{met_per}\\
G_{ij}=& D_{ij} F+ \partial_i H_j+ \partial_j H_i+ \gamma_{ij} \label{met_per2}\mbox{ ,}
\end{align}
where $D$ and $F$ are scalar perturbations, $E_i$ and $H_i$ are transverse perturbation vectors, $\partial_i D^i= \partial_i H^i=0$, $D_{ij}$ is a traceless derivative operator, $D_{ij}= \partial_i \partial_j- \frac{1}{3} \delta_{ij} \nabla^2$, and $\gamma_{ij}$ is a transverse and traceless tensor, ${\gamma_i}^i=0$, $\partial^i \gamma_{ij}=0$ (where the latin indices are raised/lowered with $\delta_{ij}$).

The inverse metric components are:
\begin{equation}\label{inverseADM}
g^{00}= -\frac{1}{N^2}, \quad g^{0i}= -\frac{N^i}{N^2}, \quad g^{ij}= h^{ij}-\frac{N^i N^j}{N^2} \mbox{ ,}
\end{equation}
where $h^{ij}$ is the inverse of $h_{ij}$ and the latin indices are lowered and raised with the metric $h_{ij}$ and its inverse (i.e. $N^i=h^{ij}N_j$).

\subsection{Gauge fixing}

\noindent Not all the perturbations just introduced are physical, but there are some gauge modes that have to be removed by gauge fixing. The gauge in which we work is the so-called \textit{spatially flat} gauge. In this gauge all the scalar perturbations of the 3-metric $h_{ij}$, $F$ and $B$, are removed leaving only the scalar perturbation of the inflaton $\delta \phi$, besides the ones in $N$ and $N_i$. Then one is free to remove also the vector perturbation $H_i$, remaining with a 3-metric $h_{ij}$ of the form :
\begin{equation}\label{gauge_fixing}
 h_{ij}= a^2[\delta_{ij}+\gamma_{ij}], \mbox{ } {\gamma_i}^i=0, \mbox{ }\partial^i \gamma_{ij}=0 \mbox{ ,}
\end{equation}
together with the scalar inflaton perturbation $\delta \phi$.

Moreover we can perform a non-linear generalizations of gauge \eqref{gauge_fixing} which is  (for more details, see, e.g, Refs.~\cite{Salopek:1990jq,Maldacena:2003, Collins:2011, Chen:2007}):
\begin{equation}\begin{split}\label{gauge_fixing_NL}
 &h_{ij}= a^2 \left[\exp \mbox{ }\gamma\right]_{ij}, \mbox{ } {\gamma_i}^i=0, \mbox{ }\partial^i \gamma_{ij}=0 \mbox{ ,}\\
&[\exp \mbox{ } \gamma]_{ij}= \delta_{ij}+ \gamma_{ij}+\frac{1}{2!}  \gamma_{ik}  {\gamma^k}_{j}+... \mbox{ .}
\end{split}\end{equation}
Now $\gamma_{ij}$ are non-linear generalizations of the variables introduced above. The advantage of this gauge choice is that in analysing primordial non-Gaussianities the terms dominant in the slow-roll hypothesis will be clear. The disadvantage is that the variable $\zeta$, the so-called \textit{curvature perturbation on comoving hyper-surfaces}, does not appear explicitly. But we can pass from the inflaton perturbation $\delta \phi$ to variable $\zeta$ through the non-linear relation (see Ref. \cite{Maldacena:2003}) 
\begin{equation}\begin{split} \label{zeta_non_linear}
\zeta & = \zeta_1 + \frac{1}{2} \frac{\ddot \phi}{\dot \phi H} \zeta_1^2 + \frac{1}{4} \frac{\dot \phi^2}{H^2} \zeta_1^2+\\ 
&+\frac{1}{H}\dot \zeta_1 \zeta_1 - \frac{1}{4} \frac{a^{-2}}{H^2} \partial^{-2} \partial_i \partial_j (\partial^i \zeta_1 \partial^j \zeta_1)+ \frac{1}{2H} \partial_i \psi \partial^i \zeta_1-\frac{1}{2 H} \partial^{-2} \partial_i \partial_j (\partial^i \psi \partial^j \zeta_1)- \frac{1}{4 H} \dot {\gamma}_{ij} \partial^i \partial^j \zeta_1 \mbox{ },
\end{split}\end{equation}
where $\phi$ is the background value of the inflaton, $H$ is the Hubble parameter, $\psi$ is as in Eq. \eqref{sol_const} and 
\begin{equation}
\zeta_1= - \frac{H}{\dot \phi} \delta \phi
\end{equation}
is the value of $\zeta$ at linear level. On large scales, much after the horizon crossing during inflation, only the first line of Eq. \eqref{zeta_non_linear} is non-vanishing (see again Ref. \cite{Maldacena:2003} for a detailed demonstration of this fact).

\subsection{Auxiliary fields}

\noindent As said previously, $N$ and $N_i$ are auxiliary fields in standard gravity, which means that they can be removed by solving the corresponding equations of motion (e.o.m.) and substituting the solutions back into the action. However, when we introduce a modified gravity term in the total Lagrangian, we have to take care of the modifications of e.o.m. of such fields. Such modifications can, e.g., make these fields dynamical, enlarging the degrees of freedom of the theory. In our case we are interested to perform an analysis of the bispectrum of primordial perturbations. In order to do this, it is necessary to expand the Lagrangian \eqref{L_studied} until third order in cosmological perturbations. As shown in Ref. \cite{Maldacena:2003}, if the fields $N$ and $N_i$ are not dynamical, they can be written in power of series of the perturbations $\delta \phi$ and $\gamma_{ij}$ in the gauge \eqref{gauge_fixing_NL}. In this case, if we are not interested in expanding the Lagrangian beyond the third order,  we can take the values of auxiliary fields just at first order in the perturbations (see. e.g.,~\cite{Maldacena:2003, Chen:2007}). In gauge \eqref{gauge_fixing_NL} and in standard gravity the solutions for auxiliary fields until first order in the perturbations are~\cite{Maldacena:2003}
\begin{equation}
N^{\delta \phi}=  1 + \frac{\dot{\phi_0}}{2H} \delta \phi \mbox{ ,} \quad  N^{\delta \phi}_i= \partial_i D , \quad D= -a^2 \frac{\dot{{\phi_0}^2}}{2 H^2} \partial^{-2} \left[ \frac{d}{dt}\left(\frac{H \delta \phi}{\dot{\phi_0}}\right)\right] \label{sol_const}\mbox{ }.
\end{equation}
Let us discuss now briefly if in our case, Eq.~\eqref{L_studied}, modifications w.r.t. standard gravity are introduced by the Chern-Simons term to the first order values of the fields $N$ and $N_i$. We notice immediately that $N$ is a scalar field and so it does not contribute to parity breaking terms, as the Chern-Simons one. In addition, $N_i$ can be splitted, Eq.~\eqref{met_per}, in a scalar perturbation mode $D$ and a transverse vector perturbation $E_i$. The Chern-Simons term can not receive any contribution from the scalar perturbation mode for the same reason as for the field $N$; instead, a priori,  it can receive a contribution from the transverse vector perturbation. In particular there are two possibilities: the first possibility is that $E_i$ remains an auxiliary field. However, at first order in the perturbations $\delta \phi$ and $\gamma_{ij}$ there is no way to build any non-trivial transverse vector. So, in this case, at first order we can take as usual $E_i = 0$ during inflation (as it happens in standard gravity, see Eq.~\eqref{sol_const}). The second possibility is that the Chern-Simons term makes the field $E_i$ dynamical: in this case we would have an additional dynamical vector perturbation that evolves during inflation. This new degree of freedom could also interact with the standard inflationary perturbations $\delta \phi$ and $\gamma_{ij}$. In any case it is not the aim of this paper to study such a possible new degree of freedom and its interactions with the fundamental fields of the standard theory.
Thus, for what we have stated, we can take for the fields $N$ and $D$ the same values as in Eq. \eqref{sol_const} and put $E_i=0$.

Now, we have all the tools we need to study the dynamical evolution of primordial perturbations.

\section{Signatures in primordial power-spectra}\label{4}

\noindent Before to start the analysis of parity breaking in primordial bispectra, in this section we briefly resume the results about parity breaking signatures in primordial power-spectra, providing also some original considerations. This is useful to clarify the conventions and the assumptions we adopt.

In the quadratic part of the Lagrangian the Chern-Simons term does not provide any contribution to the inflaton perturbation. In fact, as we have stated in the previous section, a scalar field does not perceive parity breaking signatures. Instead it is interesting to analyse the parity breaking effects into the power spectra of the transverse and traceless tensor perturbations $\gamma_{ij}$ (i.e. PGW). Let us remember the following Fourier decomposition:
\begin{equation}\label{furierdec_graviton}
\gamma_{ij}(\vec{x}, t)= \frac{1}{(2 \pi)^{3}} \int d^3k \sum_{s=s_1, s_2} \epsilon^{(s)}_{ij}(\vec{k}) \gamma_s(t, \vec k)  \mbox{ }e^{i \vec{k}\cdot \vec{x}} \mbox{ ,}
\end{equation}
where $\epsilon^{(s)}_{ij}(\vec{k})$ is the polarization tensor, $s$ is the polarization index and $\gamma_s(t, k)$ is the mode function. For our purposes it is convenient to use the so-called circular left (L) and right (R) polarization states of PGW, which are defined as:
\begin{align}\label{definition_LR}
\epsilon^R_{ij}= \frac{1}{\sqrt{2}}(\epsilon^+_{ij}+ i \epsilon^{\times}_{ij}) \mbox{ },\\
\epsilon^L_{ij}= \frac{1}{\sqrt{2}}(\epsilon^+_{ij} - i \epsilon^{\times}_{ij}) \mbox{ },
\end{align}
where $\epsilon^{\times}_{ij}$ and $\epsilon^{+}_{ij}$ are the usual two linear independent polarizations of PGW.

It is possible to show the validity of the following relations (see, e.g., Ref.~\cite{Alexander:2005}):
\begin{align}
\epsilon^L_{ij}(\vec{k})\epsilon_{L}^{ij}(\vec{k})&= \epsilon^R_{ij}(\vec{k})\epsilon_{R}^{ij}(\vec{k})= 0 \mbox{ }, \nonumber\\
\epsilon^L_{ij}(\vec{k})\epsilon_{R}^{ij}(\vec{k})&= 2 \mbox{ }, \nonumber\\
\epsilon^L_{ij}(-\vec k)&= \epsilon^R_{ij}(\vec k) \nonumber \mbox{ },\\ 
k_l \epsilon^{mlj} \mbox{ } {{\epsilon_{(s)}}_{j}}^i(\vec{k}) &= - i \mbox{ } \lambda_s k \mbox{ }\epsilon_{(s)}^{im}(\vec{k}) \nonumber \mbox{ ,}\\
\gamma^*_s(\vec k) & = \gamma_s(-\vec k) \label{relations_LR}\mbox{ ,}
\end{align}
where $k_l$ is the $l$-th component of the momentum $\vec{k}$, $\lambda_R= +1$ and $\lambda_L= -1$. Here one has to be careful not to make confusion between the Levi-Civita pseudo-tensor $\epsilon_{ijk}$ which has three latin indices and the polarization tensors $\epsilon_{ij}$ that have only two latin indices. Also we recall that $s$ is the polarization index and not a tensor index. In Eqs.~\eqref{relations_LR} latin contractions are made with $\delta_{ij}$. In addition, it is convenient to pass to the conformal time $d \tau = a^{-1} dt$ instead of adopting cosmological time $t$. The metric \eqref{ADM} reduces to the form $g_{\mu \nu}= a^2 \gamma_{\mu \nu}$, where $\gamma_{\mu \nu}$ is the perturbed Minkowski metric tensor. Since the Chern-Simons term is invariant under a Weyl transformation of the metric $g'= e^{-2 w(x,t)} g$, it is sufficient to choose $w= \ln a$ to understand that we can compute the Chern-Simons term using the metric $\gamma_{\mu \nu}$. We achieve this metric simply putting $a=1$ into the metric \eqref{ADM} and substituting cosmological time $t$ with conformal time $\tau$. 

Now, using the gauge \eqref{gauge_fixing_NL}, we compute the density Lagrangian \eqref{L_studied} at second order in tensor perturbations. Because of the fact that the Weyl tensor is zero on the background, the correction to the quadratic action from the Chern-Simons term comes from the computation of the tensor contraction $f(\phi_0)\epsilon^{\mu \nu \rho \sigma} {{C_{\mu \nu}}^{\kappa \lambda}|}^{(1)}_T {C_{\rho \sigma \kappa \lambda}|}^{(1)}_T$, where $\phi_0$ is the background value of the inflaton field. The indices $(1)$ and $T$ indicate that we need to compute the corresponding tensors at first order in tensor perturbations. For simplicity of notation, in the followings we will use $\phi$ to refer to the background value of the inflaton instead of $\phi_0$. In performing this computation we can put $N=1$ and $N_i=0$ because we are not interested in scalar perturbations. Moreover we need to use the relations \eqref{relations_LR}. The result is the total action
\begin{equation} \label{action_tt2}
S|_{\gamma \gamma}= \sum_{s=L, R} \int d\tau \mbox{ }\frac{d^3 k}{(2 \pi)^3} \mbox{  } A^2_{T, s} \left[\mbox{ }{|\gamma'_s (\tau,k)|}^2-k^2 {|\gamma_s (\tau,k)|}^2 \mbox{ }\right] \mbox{ ,}
\end{equation}
where 
\begin{equation}\label{A}
A^2_{T, s}= \frac{M^2_{Pl}}{2} a^2 \left(1 - 8 \lambda_s  \frac{k}{a} \frac{\dot f(\phi)}{M^2_{Pl}} \right) = a^2 \left(1 - \lambda_s  \frac{k_{phys}}{M_{CS}}  \right) \mbox{ }
\end{equation}
and
\begin{equation}\label{mass_CS}
M_{CS}= \frac{ M^2_{Pl}}{ 8 \dot f(\phi)} 
\end{equation}
is the so called Chern-Simons mass. Let us recall that the coefficient $\lambda_s$ is $+1$ for R polarization modes and $-1$ for L modes. So there are some values of the physical wave number $k_{phys}= k/a$ for which the factor $A^2_{T, s}$ becomes negative. In particular, from \eqref{A}, this happens for $k_{phys}> M_{CS}$. The R modes with physical wave-numbers larger than the $M_{CS}$ acquire a negative kinetic energy and become automatically ghost fields. Ghost fields at quantum level and at high energies can be very problematic (see e.g. Ref. \cite{Fulvio:2015}). In order to avoid this problem, we consider only gravitons with $k_{phys} < M_{CS}$ at the beginning of inflation. This is equivalent to put an ultraviolet cut-off $\Lambda< M_{CS}$ to the theory. At the beginning of inflation we need also gravitons with $k_{phys} \gg H$, which means that the corresponding wavelength $\lambda_{phys}$ is well inside the Hubble horizon. This is necessary to have quantum tensor perturbations that originate from a Bunch-Davies vacuum state. Thus, it follows that we have to require $M_{CS} \gg H$ during inflation in order to preserve the Bunch-Davies state of slow-roll models of inflation.

Now, instead of deriving the equations of motion for the fields $\gamma_s$, it is more convenient to make the field redefinition 
\begin{equation}\label{redefinition}
\mu_s= A_{T, s} \gamma_s \mbox{ .} 
\end{equation}
Following Ref. \cite{Satoh:2010}, the final equations of motion for the fields $\mu_s$ turn out to be
\begin{equation}\label{eom_CS2}
\mu_s''+\left(k^2-\frac{\nu_T^2-\frac{1}{4}}{\tau^2}+\lambda_s \frac{k}{\tau} \frac{H}{M_{CS}} \right)\mu_s = 0 \mbox{ ,}
\end{equation}
where $\nu_T= \frac{3}{2}+ \epsilon$, and $\epsilon$ is the slow-roll parameter
\begin{equation}
\epsilon= \frac{1}{2} \left(\frac{M_{Pl} V'}{V} \right)^2 \simeq \frac{1}{2} \frac{\dot{\phi^2}}{H^2} M^{-2}_{Pl} \label{epsilon}\mbox{ .}
\end{equation}
Eq. \eqref{eom_CS2} differs by the one of standard slow-roll models of inflation by an additional term in the effective mass proportional to $\tau^{-1}$.~\footnote{Notice also that, in the same way, the tensor sector is modified w.r.t~\cite{Maldacena:2011, Soda:2011, Shiraishi:2011} dealing with Weyl-cubic terms for which the tensor wavefunctions remain the standard ones of slow-roll inflation.} This term is dependent by the polarization index and then we have a different dynamical evolution of the two polarization states. Now, as usual, we canonically quantize the fields $ \mu_{L, R}$ as
\begin{equation}
\hat \mu_s (k, \tau) =  u_s(k, \tau) \hat{a}_s(\vec k) +  u^*_s(k, \tau) \hat{a}^{\dagger}_s(- \vec k)  \mbox{ },
\end{equation}
where the creation and annihilation operators obey the usual relations
\begin{align}
 \langle0|\hat{a_s}^{\dagger}&=0 , \quad \hat a_s |0\rangle =0 \label{canonical_quant} ,\\
[\hat a_s (k), \hat{a}^{\dagger}_{s'} (k')]&= (2\pi)^3 \delta^3(k-k')  \delta_{s s'}, \quad [\hat a_{k}, \hat{a}_{k'}]= [\hat a_{k}^{\dagger}, \hat{a}^{\dagger}_{k'}]= 0 \label{canonical_quant2}\mbox{ }.
\end{align}
Thus, the equations of motion for the mode functions $u_s(k, \tau)$ are 

\begin{equation}\label{eomCS_3}
u_s''+\left(k^2-\frac{\nu_T^2-\frac{1}{4}}{\tau^2}+\lambda_s \frac{k}{\tau} \frac{H}{M_{CS}}  \right)u_s = 0 \mbox{ .}
\end{equation}
The exact solution of Eq. \eqref{eomCS_3} with the Bunch Davies initial condition reads like (see Ref. \cite{Satoh:2010})
\begin{equation} \label{solution_CS}
 u_s(k, \tau)= 2 \sqrt{\frac{(-k \tau)^{3}}{k}} e^{-i(\frac{\pi}{4}- \pi \nu_T/2)} \mbox{ } e^{-i k \tau} \mbox{ } U \left(\frac{1}{2}+ \nu_T - \lambda_s \frac{H}{M_{CS}}, 1+ 2 \nu_T, 2i k \tau \right) \mbox{ }e^{+\frac{\pi}{4} \lambda_s H/M_{CS}} \mbox{ },
\end{equation}
where $U$ is the confluent hypergeometric function \cite{NIST:DLMF}. On super-horizon scales (i.e. $z=-k \tau \ll 1$) the solution \eqref{solution_CS} simplifies, becoming 
\begin{equation}
u_s(k, \tau)_{z \ll1}= \sqrt{\frac{1}{2 k^2 \tau^2 k}} e^{i(-\frac{\pi}{4}+ \frac{\pi}{2} \nu_T)} \frac{\Gamma(\nu_T)}{\Gamma(3/2)} \left( \frac{-k \tau}{2}\right)^{3-2 \nu_T} \mbox{ }e^{+\frac{\pi}{4}\lambda_s H/M_{CS}} \mbox{ }.
\end{equation}
Now, we can compute the super-horizon power spectra of each polarization mode.
In the quantum field theory statistical ensamble they read
\begin{equation}
\Delta_T^L= \langle 0 \vert \hat{\gamma}_{ij}^L(\vec k) \hat{\gamma}^{ij}_L(\vec k')\vert 0\rangle = 2 \frac{| u_L(z)_{z \ll1}|^2}{A^2_{T,L}} \mbox{ },
\end{equation}

\begin{equation}
\Delta_T^R= \langle 0 \vert \hat{\gamma}_{ij}^R(\vec k) \hat{\gamma}^{ij}_R(\vec k') \vert 0\rangle = 2 \frac{| u_R(z)_{z \ll1}|^2}{A^2_{T,R}} \mbox{ }.
\end{equation}
 At leading order in the slow-roll parameters we find \cite{Satoh:2010}
\begin{eqnarray}\label{delta_expansion}
\Delta_T^L= & \frac{\Delta_T}{2} e^{ -\frac{\pi}{4}  H/M_{CS}} \mbox{ }, \\
\Delta_T^R= &\frac{\Delta_T}{2} e^{ + \frac{\pi}{4}  H/M_{CS}} \mbox{ } ,
\end{eqnarray}
where 
\begin{equation} \label{power_spectrum_tensor}
\Delta_T = \frac{4}{k^3} \frac{H^2}{M^2_{Pl}} \left( \frac{z}{2}\right)^{3-2 \nu_T} \mbox{ }.
\end{equation}
Notice that $\Delta_T$ is the total power spectrum of tensor perturbations in the standard slow-roll models without the Chern-Simons correction. As we explained before the dimensionless coefficient $H/M_{CS}$ is assumed to be much smaller than 1. For this reason we can expand in series the exponentials. At the end, the relative difference between the power spectrum of right ($R$) and left ($L$) helicity states reads~\cite{Satoh:2010}
\begin{equation}\label{theta}
\Theta = \frac{\Delta_T^R - \Delta_T^L}{\Delta_T^R+ \Delta_T^L} = \frac{\pi}{2} \left( \frac{H}{{M}_{CS}} \right)\mbox{ }.
\end{equation}
This observable quantifies the differences between the power spectrum of the helicity polarizations L and R of the primordial gravitational waves. We expect that its value is small (i.e. $\ll1$) for the considerations made in developing the theory. In the literature this type of parity violating signature in the power spectrum of the primordial gravitational waves was firstly considered in Ref. \cite{Lue:1999}. 

As we mentioned in the Introduction, even in the most optimistic cases, only weak constraints on the parameter~\eqref{theta} could be achieved from future CMB power spectra (see e.g. Refs. \cite{Gerbino:2016, Planck_parity_breaking_2016}). In the future one expects to improve the constraints on the value of $\Theta$ with future experiments involving the detection of polarized primordial gravitational waves through interferometers (see e.g. Refs. \cite{Seto:2007, Seto:2008, Crowder:2013}). 

Another interesting feature of the model is that we can use \eqref{delta_expansion} to compute the modifications to the tensor-to-scalar-ratio of the standard theory. In fact the new total dimensionless power spectrum of tensor perturbations reads in terms of $\Theta$ as 
\begin{equation}
\Delta^{CS}_T= \Delta_T^R+ \Delta_T^L= \Delta_T \left[1+\frac{\pi^2}{16} \left(\frac{H}{M_{CS}}\right)^2  \right]= \Delta_T \left( 1+ \frac{\Theta^2}{4}\right) \mbox{ }.
\end{equation} 
On the contrary, the dimensionless scalar power spectrum $\Delta_S$ does not receive any contribution due to parity symmetry of the scalar perturbations. Thus we find for the tensor-to-scalar ratio
\begin{equation}\label{r_CS}
r_{C-S}= \frac{\Delta^{C-S}_T}{\Delta_S}=  \frac{\Delta_T}{\Delta_S} \left( 1+ \frac{\Theta^2}{4} \right)= r \left( 1+ \frac{\Theta^2}{4}\right) \mbox{ },  
\end{equation}
where $r$ is the tensor-to-scalar-ratio of the slow-roll models without the Chern-Simons term. In addition, it is easy to show that, as a first approximation,~\footnote{It turns out that 
\begin{equation}
n_T=-2 \epsilon+ \pi^2 \left[ 2 H^2 f''(\phi) \epsilon-\frac{1}{8} \frac{H}{M_{CS}}  \eta  \right] \left( \frac{H}{M_{CS}} \right) \left[ 1-\frac{\pi^2}{16} \left( \frac{H}{M_{CS}} \right)^2 \right]\, .
\end{equation}} the spectral index of tensor perturbations $n_T$ remains the same of the standard slow-roll models of inflation (at leading order in the slow-roll parameters and taking into account that $(H/M_{CS}) \ll 1$). Therefore a modification to the usual consistency relation follows
\begin{equation}\label{modification_consistency}
r_{C-S}\simeq -8 n_T \left(1+ \frac{\Theta^2}{4}\right) \mbox{ } ,
\end{equation}
where we have used Eq. \eqref{r_CS} and the standard consistency relation of slow-roll models which is given by
\begin{equation}\label{consistency}
r = -8 n_T \mbox{ } .
\end{equation}
 We remind that the tensor-to-scalar ratio is not sensitive to the polarizations of the gravitational waves, because it refers to the total power spectra. Thus a priori, using \eqref{modification_consistency}, one could measure the effects of the Chern-Simons term also searching for unpolarized primordial gravitational waves. But in this case the corrections are of order $\Theta^2$ and probably this effect is even more difficult to observe.

Briefly speaking, in the model we have considered the formation of a very small parity violation in the PGW power spectrum is expected (see Eq. \eqref{theta}). This might be very difficult to observe. For this reason it is interesting to investigate if a significant parity breaking signature can arise from a higher order statistics of primordial perturbations. In the following section we will study how the Chern-Simons term coupled to the inflaton affects the bispectra of primordial perturbations.

\section{Signatures in primordial bispectra}\label{5}

\noindent 
In this section we are interested to understand whether the effects of the new interaction terms coming from the Chern-Simons term can bring to a relevant parity breaking effect in the bispectra of the primordial perturbations. 
In the subsequent three subsections we will estimate the primordial  three-point functions $\langle \gamma \gamma \gamma \rangle$ and $\langle \gamma \zeta \zeta \rangle$ and compute explicitly $\langle \gamma \gamma \zeta \rangle$. In fact, as we have anticipated in the introduction, we will show that the parity breaking contribution to the correlator $\langle \gamma \gamma \zeta \rangle$ is the only one that is not suppressed as the power spectrum case. The readers who are not interested in the technical details of the computation of such three-point functions can go directly to subsection \ref{5.4} where we will provide our main results, including the explicit expression of $\langle \gamma \gamma \zeta \rangle$, and a discussion on its parity breaking signatures.

For parity invariance reasons  the only primordial bispectra that can arise from the Chern-Simons term are the ones associated to the three-point functions $\langle \gamma \gamma \gamma \rangle$, $\langle \gamma \gamma \delta \phi \rangle$ and $\langle \gamma \delta \phi \delta \phi\rangle$. We recall that, given three generic cosmological perturbations $\delta_{a}(\vec x,t)$, $\delta_{b}(\vec x,t)$, $\delta_{c}(\vec x,t)$, the bispectrum $B(k_1, k_2, k_3)$ associated to the three-point function $\langle \delta_{a}(\vec x_1, t) \delta_{b} (\vec x_2, t) \delta_{c}(\vec{x_3}, t) \rangle$ is its Fourier transform. Thus, in formula
\begin{equation}
\langle \delta_{a}(\vec{k_1}) \delta_{b}(\vec k_2) \delta_{c}(\vec k_3) \rangle = (2 \pi)^3 \delta^{(3)}({\vec k_1+ \vec k_2 + \vec k_3}) B(k_1, k_2, k_3) \mbox{ ,}
\end{equation}
where the Dirac delta is due to invariance under translations and $B(k_1, k_2, k_3)$ depends only on the wave numbers $k_i$ due to invariance under rotations. In the following sections sometimes we will forget about the Dirac delta, for simplicity of notation.

The master formula used to compute this kind of correlators in the quantum field theory ensamble is provided by the in-in formalism (see e.g. Refs. \cite{Weinberg:2005,Maldacena:2003, Collins:2011, Chen:2007})
\begin{equation}\label{master}
\langle \delta_{a}(\vec{k_1}) \delta_{b}(\vec k_2) \delta_{c}(\vec k_3) \rangle = - i \int_{t_0}^t dt' \langle 0 \vert \Big[\delta_{a}(\vec{k_1},t)\delta_{b}(\vec{k_2},t) \delta_{c}(\vec{k_3},t)\mbox{ }, \mbox{ }H_{int}(t')\Big]  \vert 0\rangle \mbox{ }, 
\end{equation}
where $H_{int} = - L_{int}$ is the cubic interaction Hamiltonian between fields $\delta_{a}$, $\delta_{b}$ and $\delta_{c}$,  $t_0$ is the time at which this interaction is switched on and $t$ is the time at which we evaluate the correlator. In the formula just introduced $t$ is a totally arbitrary time coordinate; it will be very convenient for the computations to adopt the conformal time $\tau$ instead of the cosmological time. The interactions are switched on on sub-horizon scales, that correspond to the limit $\tau_0 = -\infty$; on the other hand we evaluate the correlator on super-horizon scales, then in the limit $\tau = 0$. Now, let us start to evaluate the interaction Lagrangians for the bispectra in which we are interested in.
\subsection{Three gravitons non-Gaussianities}

\noindent  This interaction comes from the following contribution to the Lagrangian
\begin{equation}\label{ttt_first}
\mathcal{L}_{int}^{\gamma \gamma \gamma} = f(\phi)\epsilon^{\mu\nu\rho\sigma} \Big[ C_{\mu\nu}{}^{\kappa\lambda}|^{(2)}_{T}C_{\rho\sigma\kappa\lambda}|_T^{(1)}+ C_{\mu\nu}{}^{\kappa\lambda}|^{(1)}_{T}C_{\rho\sigma\kappa\lambda}|_T^{(2)} \Big]\, ,
\end{equation}
where the suffix $T$ indicates that we have to evaluate the contribution of the tensor perturbations only and the index $(n)$ means that we have to consider the n-th order of the perturbations contributing to the corresponding Weyl tensor. We will use an identical notation also for the other interaction vertices.
As we can see, the explicit computation of this term requires to compute the Weyl tensor up to second order in the tensor perturbations. Instead of performing a direct computation, we can have an idea of the strength of this interaction vertex using only the tensorial properties of the Chern-Simons term.  In fact, as we have recalled in Sec. 2, the Chern-Simons term is a total derivative term for a constant $f$, so that we can write \begin{equation}\label{L_ttt}
\mathcal{L}_{int}^{\gamma \gamma \gamma} = f(\phi) A^{\mu}_{\,\,; \mu}\, ,
\end{equation}
where $A^{\mu}$ is a four-vector, whose expression can be found, e.g., in Refs.~\cite{Grumiller:2007rv,Jackiw:2003} and the semi-column stands for covariant derivative.~\footnote{With the notation used in~(\ref{L_ttt}) we indicate that we will take the part of the Chern-Simons interaction that gives rise to a graviton cubic interaction $\langle \gamma \gamma \gamma \rangle$.} However even without considering its explicit expression, we can understand what is the general form of $A^{\mu}$. As we said in Sec. 3, we can compute the Weyl tensor using the metric \eqref{ADM}, using conformal time instead of cosmological time and setting the scale factor $a=1$. The result is that $A^{\mu}$ has to depend only by $\gamma_{ij}$ and derivatives of $\gamma_{ij}$ and by no additional factors $a$ or $H$. Thus $A^{\mu}$ is a cubic combination of tensor perturbations and their derivatives only. In particular, integrating by parts the Lagrangian \eqref{L_ttt} it follows
\begin{equation}\label{L_ttt2}
\mathcal{L}_{int}^{\gamma \gamma \gamma} = - f'(\phi)  A^{0},
\end{equation}
 where, for our purposes, we can just evaluate the coupling $f(\phi)$ on the background. Here the prime $'$ refers to the derivative w.r.t. conformal time. If we put this interaction into formula \eqref{master} we find
\begin{equation}\label{master_graviton}
\langle \gamma_{s_1}(\vec{k_1}) \gamma_{s_2}(\vec k_2) \gamma_{s_3}(\vec k_3) \rangle = i \int_{-\infty}^0 d\tau' a \dot f(\phi)\langle 0 \vert \Big[\gamma_{s_1}(\vec{k_1}, 0)\gamma_{s_2}(\vec{k_2}, 0) \gamma_{s_3}(\vec{k_3}, 0)\mbox{ }, \mbox{ }A^{0}\Big]  \vert 0\rangle \mbox{ },
\end{equation}
where we have used the fact that $f'(\phi)= a \dot f(\phi)$.

The bra-ket contractions into Eq.~\eqref{master_graviton} produces products of three Green functions of the type
\begin{equation}
\langle \gamma_{s_1}(\vec{p_1}, 0) \gamma_{s_2}(\vec p_2, \tau') \rangle = (2 \pi)^3 \delta^{(3)}(\vec p_1 + \vec p_2) u_{s_1}(p_1, 0) u_{s_2}(p_2, \tau'),
\end{equation}
with $u_{s}(k, \tau)$ given as in Eq. \eqref{solution_CS}, apart for the contribution of field redefinition \eqref{redefinition}. As a first approximation we can use the expression of $u_s$ taking the slow-roll parameters and $H/M_{CS}$ to be vanishing. This is justified by the fact that, under our assumptions, such parameters are much smaller than $1$ during inflation and thus we can expand in series the solution \eqref{solution_CS} and take the zero-th order value. The solution becomes 
\begin{equation}\label{u_integrable}
u_s(k, \tau) = \frac{i H}{M_{Pl} \sqrt{k^3}} (1+ik \tau) e^{-i k \tau} \mbox{ ,}
\end{equation}
which is the solution for the mode function of a scalar field in a de Sitter space. 

Then, substituting \eqref{u_integrable} into Eq. \eqref{master_graviton} we will arrive to an integral of the type
\begin{equation}\label{master_graviton2}
\langle \gamma_{s_1}(\vec{k_1}) \gamma_{s_2}(\vec k_2) \gamma_{s_3}(\vec k_3) \rangle = i \int_{-\infty}^0 d\tau' \left(-\frac{1}{H \tau}\right) \dot f(\phi)  \left(\prod_i \frac{H^2}{M^2_{Pl} k_i^3} \right) f(k_i, \tau') e^{-ik_T \tau'}  \mbox{ },
\end{equation}
where $k_T= k_1+k_2+k_3$, and $f(k_i, \tau')$ is a function polynomial in the arguments. We have used $a= -1/(H \tau)$, a relation which holds apart for small slow-roll corrections. Due to the slow-roll dynamics we can treat $H$ and $\dot f(\phi)$ as approximately constant parameters and put them out of the integrals. In particular we can minimize the errors committed taking the value of such parameters near the horizon crossing of the overall momenta $K= K_T$. In fact after and much before horizon crossing the integral in Eq. \eqref{master_graviton2} vanishes (see Ref. \cite{Maldacena:2003}  and also, e.g.,~\cite{Collins:2011} for more details). We know that much before horizon crossing the integrand function has a high oscillatory behaviour because of the imaginary exponential, which mediates the integral to zero; after horizon crossing the tensor perturbations $\gamma_{ij}$ become constant and then the commutator operator with the interaction Lagrangian in Eq. \eqref{master_graviton} becomes zero. Thus, we have
\begin{equation}\label{master_graviton3}
\langle \gamma_{s_1}(\vec{k_1}) \gamma_{s_2}(\vec k_2) \gamma_{s_3}(\vec k_3) \rangle = i \dot f_*(\phi)  \left(\prod_i \frac{H_*^2}{M^2_{Pl} k_i^3}\right) \frac{1}{H_*}  \int_{-\infty}^0 d\tau' \left(-\frac{1}{\tau}\right) f(k_i, \tau') e^{-ik_T \tau'}   \mbox{ },
\end{equation}
where the $*$ refers to the time of horizon crossing of the overall momenta. In the following we will omit the $*$ for simplicity of notation. Integrals of the type in Eq. \eqref{master_graviton3} can be computed by parts after passing to complex plane and performing a Wick rotation (see e.g. Refs. \cite{Maldacena:2003, Collins:2011, Chen:2007}). The result is an imaginary function which depends on the momenta $k_i$ and has the physical dimensions of a $(mass)^3$. In addition from Eq. \eqref{mass_CS} , it follows
\begin{equation}
\dot f(\phi) = \frac{M^2_{Pl}}{8 M_{CS}} \mbox{ .}
\end{equation}
Thus, if we multiply and divide Eq. \eqref{master_graviton3} by $\sum_i k_i^3$, we find out the following estimation:
\begin{equation}\label{master_graviton4}
\langle \gamma_{s_1}(\vec{k_1}) \gamma_{s_2}(\vec k_2) \gamma_{s_3}(\vec k_3) \rangle \sim  \frac{H}{M_{CS}} \left(\sum_{i\neq j} \Delta_T(k_i) \Delta_T(k_j)\right) M(k_i),
\end{equation}
$M(k_i)$ being a dimensionless function of the momenta $k_i$ and $\Delta_T(k)$ the power spectrum of tensor perturbations, Eq. \eqref{power_spectrum_tensor}. Due to momentum conservation the function $M(k_i)$ should be of order $1$ and gives only the momentum shape of the bispectrum. As we can see from this final result, parity breaking in graviton non-Gaussianities is suppressed by the ratio $H/M_{CS}$, exactly as the power spectra case.

\subsection{Two scalars and a graviton non-Gaussianities}

\noindent Here we are interested in the following interaction Lagrangian
\begin{equation}\label{L_sst}
\begin{split}
\mathcal{L}_{int}^{\delta \phi \delta \phi \gamma} = \epsilon^{\mu\nu\rho\sigma} &\Big[\frac{\partial f(\phi)}{\partial \phi} \delta \phi \mbox{ } C_{\mu\nu}{}^{\kappa\lambda}|^{(1)}_{S}C_{\rho\sigma\kappa\lambda}|_T^{(1)}+ \frac{\partial f(\phi)}{\partial \phi} \delta \phi \mbox{ }C_{\mu\nu}{}^{\kappa\lambda}|^{(1)}_{T} C_{\rho\sigma\kappa\lambda}|_S^{(1)} + f(\phi) C_{\mu\nu}{}^{\kappa\lambda}|^{(2)}_{S}C_{\rho\sigma\kappa\lambda}|_T^{(1)}\\
&+ f(\phi) C_{\mu\nu}{}^{\kappa\lambda}|^{(1)}_{T}C_{\rho\sigma\kappa\lambda}|_S^{(2)}+f(\phi) C_{\mu\nu}{}^{\kappa\lambda}|^{(2)}_{ST}C_{\rho\sigma\kappa\lambda}|_S^{(1)}+ f(\phi) C_{\mu\nu}{}^{\kappa\lambda}|^{(1)}_{S}C_{\rho\sigma\kappa\lambda}|_{ST}^{(2)} \Big] \mbox{ ,}
\end{split}
\end{equation}
where the suffix $S$ means that we have to evaluate the contribution only of the scalar perturbations to the corresponding Weyl tensor, and the double suffix $ST$ means that we are evaluating the quadratic contribution in which both scalar and tensor perturbations appear.

In Eq. \eqref{L_sst} the first two terms come from the expansion in series of the function $f(\phi)$ around the background value of the inflaton multiplied by the contraction of two Weyl tensors at first order in tensor perturbations; the other terms instead come from the contraction between the Weyl tensor at second order in scalar perturbations and the Weyl tensor at first order in tensor perturbations. Finally there are also contributions coming from the contraction between the Weyl tensor at second order sourced by mixed scalar and tensor perturbations and the Weyl tensor at first order in scalar perturbations. In particular, in the spatially flat gauge, scalar perturbations appear only through first order expressions of the fields $N$ and $N_i$; if we take these explicit expressions (see Eq. \eqref{sol_const}) and we use the definition of slow-roll parameter $\epsilon$ \eqref{epsilon}, it follows
\begin{equation}
N \sim \sqrt{\epsilon} \delta \phi, \qquad N_i \sim \sqrt{\epsilon} \delta \phi \mbox{ .}
\end{equation}
Thus $N$ and $N_i$ are sub-dominant in the slow-roll hypothesis in comparison with the inflaton perturbation $\delta \phi$ . For this reason in the slow-roll limit the terms dominant in Eq. \eqref{L_sst} are the first two. These terms depend only by the Weyl tensor at first order and can be easily computed. We obtain
\begin{equation}\label{L_sst2}
\mathcal{L}_{int}^{\delta \phi \delta \phi \gamma}= - 8 \frac{\partial f(\phi)}{\partial \phi} \sqrt{\epsilon}  \mbox{ } (\partial^l \delta \phi) \epsilon^{ijk} \Big[ (\partial_k \delta \phi) \partial_i \gamma'_{lj} \Big] \mbox{ .}
\end{equation}
Using again~\eqref{mass_CS}, we can express 
\begin{equation}\label{relation_partial}
\frac{\partial f(\phi)}{\partial \phi}= \frac{M^2_{Pl}}{8 M_{CS} \dot \phi} \mbox{ ,}
\end{equation}
and inserting this expression into Eq. \eqref{L_sst2} we find
\begin{equation}\label{L_sst3}
\mathcal{L}_{int}^{\delta \phi \delta \phi \gamma}= - \frac{M^2_{Pl}}{M_{CS} \dot \phi} \sqrt{\epsilon} \mbox{ } (\partial^l \delta \phi) \epsilon^{ijk} \Big[ (\partial_k \delta \phi) \partial_i \gamma'_{lj} \Big] \mbox{ .}
\end{equation}
If we use \eqref{master}, we can perform an estimation similar for what we have done for the three-graviton non-Gaussianities. The result is
\begin{equation}\label{master_sst}
\langle \delta \phi(\vec k_1) \delta \phi(\vec k_2) \gamma_{s}(\vec k_3) \rangle \sim  \frac{H}{M_{CS}} \left(\sum_{i\neq j} \Delta_T(k_i) \Delta_T(k_j)\right) F(k_i)\mbox{ },
\end{equation}
where $F(k_i)$ is another dimensionless function of order $1$. The result is that parity breaking signature in such a bipectrum is still suppressed by the ratio $H/M_{CS}$.

\subsection{One scalar and two graviton non-Gaussianities}

\noindent The interaction Lagrangian which gives contributions to this bispectrum is
\begin{equation}\begin{split}\label{L_stt}
\mathcal{L}_{int}^{\delta \phi \gamma \gamma}= &\mbox{ } \epsilon^{\mu \nu \rho \sigma} \Big[\left(\frac{\partial}{\partial \phi} f(\phi)\right) \delta \phi \mbox{ } {C^{(1)}_{\mu \nu}}^{\kappa \lambda}|_T C^{(1)}_{\rho \sigma \kappa \lambda}|_T + f(\phi)  {C^{(1)}_{\mu \nu}}^{\kappa \lambda}|_S C^{(2)}_{\rho \sigma \kappa \lambda}|_T+ f(\phi) {C^{(2)}_{\mu \nu}}^{\kappa \lambda}|_T C^{(1)}_{\rho \sigma \kappa \lambda}|_S  \\
& + f(\phi)  {C^{(1)}_{\mu \nu}}^{\kappa \lambda}|_T C^{(2)}_{\rho \sigma \kappa \lambda}|_{ST}+  f(\phi) {C^{(2)}_{\mu \nu}}^{\kappa \lambda}|_{ST} C^{(1)}_{\rho \sigma \kappa \lambda}|_{T} \Big] \mbox{ ,}
\end{split}
\end{equation}
where the notations are the same of Eqs. \eqref{ttt_first}, \eqref{L_sst}. 

Here the first term comes from the expansion in series of the function $f(\phi)$ around the background value of the inflaton multiplied by the contraction of two Weyl tensors at first order in tensor perturbations; then there are terms coming from the contraction between the Weyl tensor at second order in tensor perturbations and the Weyl tensor at first order in scalar perturbations. Finally there are some ``mixed'' terms. In this case contributions coming from the contraction between the Weyl tensor at second order sourced by mixed scalar and tensor perturbations and the Weyl tensor at first order in tensor perturbations appear. In analogy with the previous case, in the slow-roll hypothesis the term dominant in Eq. \eqref{L_stt} is the first one. In this case we obtain the cubic Lagrangian in Fourier space:
\begin{equation}\begin{split} \label{L_stt3}
 L_{int}^{\delta \phi \gamma \gamma}(\tau) =  - \lambda_s \times \int d^3K & \frac{\delta^3(\vec k+ \vec p+ \vec q)}{(2 \pi)^6}  \left\{  \left(\frac{\partial f(\phi)}{\partial \phi} \right) p \mbox{ } \delta \phi'(\vec k) \left[{\gamma'}^s_{ij}(\vec p) {\gamma'}_s^{ij}(\vec q) + \left( \vec{p} \cdot \vec{q} \right)\gamma^s_{ij} (\vec p) \gamma_s^{ij} (\vec q)\right] + \right. \\
 & \left. + a \left(\dot \phi \frac{\partial^2 f(\phi)}{\partial^2 \phi} \right) p \mbox{ }\delta \phi(\vec k) \left[{\gamma'}^s_{ij}(\vec p) {\gamma'}_s^{ij}(\vec q) +  \left( \vec{p} \cdot \vec{q} \right)\gamma^s_{ij} (\vec p) \gamma_s^{ij} (\vec q)  \right] \right\}\mbox{ },\end{split}
\end{equation}
where we have used the notation $\int d^3k \mbox{ }d^3p \mbox{ }d^3q= \int d^3 K$ and a sum over the polarization index $s$ is understood for simplicity of notation.

In this Lagrangian we see that there are some interaction vertices that depend on the second derivative of the coupling $f(\phi)$ w.r.t. the inflaton field. These interactions might give a contribution which is not suppressed directly by the ratio $H/M_{CS}$, as we have seen in the previous cases. In fact in those cases we have dealt with the first derivative of the coupling function $f(\phi)$ w.r.t. the inflaton. For this reason we perform now a detailed computation, using the in-in formalism in order to show that these new interaction vertices can bring a potentially relevant parity breaking signature to the bispectra we are analysing. At the end we will quantify if this signature is still highly suppressed or not.
 
So, we are going to use the general formula \eqref{master} to compute the correlator
\begin{equation}\label{correlator_stt}
\langle \gamma_{s_1}(\vec{k_1}, 0) \gamma_{s_2}(\vec{k_2}, 0) \delta \phi(\vec{k_3}, 0) \rangle \mbox{ }.
\end{equation}
In the next steps we will omit the time argument $\tau=0$ again for simplicity of notation. By the form of the interaction Lagrangian \eqref{L_stt3}, it follows that the only non-vanishing correlators of the type \eqref{correlator_stt} are
\begin{equation}
\langle  \gamma_{R}(\vec{k_1}) \gamma_{R}(\vec{k_2}) \delta \phi(\vec{k_3}) \rangle  ,\qquad \langle  \gamma_{L}(\vec{k_1})\gamma_{L}(\vec{k_2}) \delta \phi(\vec{k_3}) \rangle \mbox{ }.
\end{equation}
In fact we have explicitly checked that
\begin{equation}
\langle \gamma_{R}(\vec{k_1}) \gamma_{L}(\vec{k_2}) \delta \phi(\vec{k_3}) \rangle = 0 \mbox{ }.
\end{equation}
We start from the computation of the correlator $\langle \gamma_{R}(\vec{k_1}) \gamma_{R}(\vec{k_2}) \delta \phi(\vec{k_3}) \rangle$. The computations for the other correlator will be analogous.

Using Eq. \eqref{master}, we have
\begin{equation}\begin{split} \label{correlator_stt2}
\langle \gamma_{R}(\vec k_1)  \gamma_{R}(\vec k_2) \delta \phi(\vec k_3) \rangle = - \frac{i}{(2 \pi)^6} \int d^3K \mbox{ }\delta^3(\vec k+ \vec p+\vec q) \mbox{ } \int_{- \infty}^0 d\tau' & \left[ \frac{\partial f(\phi)}{\partial \phi}  \Big(\mathcal{B}_1(\tau') + \mathcal{B}_2(\tau') \Big) +\right.\\
&\left.+ a \mbox{ }\dot \phi \frac{\partial^2 f(\phi)}{\partial^2 \phi} \Big(\mathcal{B}_3(\tau') +   \mathcal{B}_4(\tau')\Big) \right]\mbox{ },\end{split} 
\end{equation}
where
\begin{align}
\mathcal{B}_1= &p \langle 0 \vert \mbox{ } \Big[\delta \phi(\vec{k_1}, 0) \gamma_R(\vec{k_2}, 0) \gamma_R(\vec{k_3}, 0) \mbox{ }, \mbox{ } \delta \phi'( \vec k, \tau') {\gamma'}^R_{ij}(\vec p, \tau') {\gamma'}_R^{ij}(\vec q, \tau')\Big]  \mbox{ }\vert 0\rangle ,\\
\mathcal{B}_2= &p \left( \vec{p} \cdot \vec{q} \right) \langle 0 \vert \mbox{ } \Big[\delta \phi(\vec{k_1}, 0)  \gamma_R(\vec{k_2}, 0) \gamma_R(\vec{k_3}, 0) \mbox{ }, \mbox{ }  \delta \phi'(\vec k, \tau') {\gamma}^R_{ij}(\vec p, \tau') {\gamma}_R^{ij}(\vec q, \tau')\Big]  \mbox{ }\vert 0\rangle ,\\
\mathcal{B}_3= &p \langle 0 \vert \mbox{ } \Big[\delta \phi(\vec{k_1}, 0) \gamma_R(\vec{k_2}, 0) \gamma_R(\vec{k_3}, 0) \mbox{ }, \mbox{ } \delta \phi(\vec k, \tau') {\gamma'}^R_{ij}(\vec p, \tau') {\gamma'}_R^{ij}(\vec q, \tau')\Big]  \mbox{ }\vert 0\rangle ,\\
\mathcal{B}_4= &p  \left( \vec{p} \cdot \vec{q} \right)\langle 0 \vert \mbox{ } \Big[\delta \phi(\vec{k_1}, 0)  \gamma_R(\vec{k_2}, 0) \gamma_R(\vec{k_3}, 0) \mbox{ }, \mbox{ } \delta \phi(\vec k, \tau') {\gamma}^R_{ij}(\vec p, \tau') { \gamma}_R^{ij}(\vec q, \tau')\Big]  \mbox{ }\vert 0\rangle .
\end{align}
Here the symbol $\Big[ \cdot\mbox{ }, \mbox{ }\cdot \Big]$ denotes the commutator operator. We can compute these expressions by using the Wick theorem. We need to compute preliminary the following contractions between fields:
\begin{align}
&\langle 0 \vert \delta \phi(\vec k, \tau) \delta \phi(\vec k', \tau')\vert 0 \rangle = (2 \pi)^3 \delta^3(\vec k+\vec k')u(k,\tau)u^*(k,\tau') ,\\
&\langle 0 \vert \delta \phi(\vec k, \tau) \delta \phi'(\vec k', \tau')  \vert 0 \rangle= (2 \pi)^3 \delta^3(\vec k+ \vec k')u(k,\tau) \frac{d}{d \tau}u^*(k,\tau')  , \\
&\langle 0 \vert \gamma^R_{ij}(\vec k, \tau) \gamma_R(\vec k', \tau')\vert 0 \rangle = (2 \pi)^3 \delta^3(\vec k+\vec k') u_R(k,\tau) u_R^*(k,\tau') \epsilon^R_{ij}(\vec k) , \\
&\langle 0 \vert  \gamma^R(\vec k, \tau) \gamma^R_{ij}(\vec k', \tau')\vert 0 \rangle = (2 \pi)^3 \delta^3(\vec k+\vec k') u_R(k,\tau) u_R^*(k,\tau') {\epsilon^R_{ij}}^*(\vec k) , \\
&\langle 0 \vert \gamma^R_{ij}(\vec k, \tau) {\gamma}'^R(\vec k', \tau')\vert 0 \rangle = (2 \pi)^3 \delta^3(\vec k+\vec k')  u_R(k,\tau)\left(\frac{d u_R^*(k,\tau')}{d \tau}  \right) \epsilon^R_{ij}(\vec k) ,\\
&\langle 0 \vert {\gamma}'^R(\vec k, \tau) \gamma^R_{ij}(\vec k', \tau')\vert 0 \rangle = (2 \pi)^3 \delta^3(\vec k+ \vec k') \left( \frac{d u_R(k,\tau)}{d \tau} \right) u_R^*(k,\tau') {\epsilon^R_{ij}}^*(\vec k) ,
\end{align}
where $u(k, \tau)$ is the mode function of the inflaton perturbation and $u_s(k, \tau)$ is the one of the tensor perturbations.
Thus, performing all the contractions, Eq. \eqref{correlator_stt2} becomes
\begin{equation}\begin{split} \label{correlator_stt3}
\langle \gamma_{R}(\vec k_1) \gamma_{R}(\vec k_2) \delta \phi(\vec k_3) \rangle =& - i (2 \pi)^3 \delta^3(k_1+k_2+k_3) \mbox{ } \operatorname{Im}\{ [ k_1 (I_1 + I_2)+ k_1 ( \vec k_1 \cdot \vec k_2) (I_3 + I_4) ] \times \\
& \times {\epsilon^R_{ij}}^*(\vec k_1) {\epsilon_R^{ij}}^*(\vec k_2) -c.c. \}+ (\vec k_1 \longleftrightarrow \vec k_2) \mbox{ },
\end{split} \end{equation}
where the $I_n$ are the integrals
\begin{align}
I_1 &= u(k_1, 0) u_R(k_1, 0) u_R(k_2, 0)\int_{- \infty}^0 d\tau' \left(\frac{\partial}{\partial \phi}  f(\phi) \right) \left[ \frac{d}{d \tau}u^*(k_1,\tau') \frac{d}{d \tau}u_R^*(k_1,\tau') \frac{d}{d \tau}u_R^*(k_2,\tau')\right] ,\\
I_2 &= u(k_1, 0) u_R(k_2, 0) u_R(k_3, 0)\int_{- \infty}^0 d\tau' a \left(\frac{\partial}{\partial \phi}  \dot f(\phi) \right) \left[ u^*(k_1,\tau') \frac{d}{d \tau}u_R^*(k_2,\tau') \frac{d}{d \tau}u_R^*(k_3,\tau')\right] ,\\
I_3 &= u(k_1, 0) u_R(k_2, 0) u_R(k_3, 0)\int_{- \infty}^0 d\tau'  \left(\frac{\partial}{\partial \phi}   f(\phi) \right) \left[ \frac{d}{d \tau}u^*(k_1,\tau') u_R^*(k_2,\tau') u_R^*(k_3,\tau')\right] ,\\
I_4 &= u(k_1, 0) u_R(k_2, 0) u_R(k_3, 0)\int_{- \infty}^0 d\tau' a \left(\frac{\partial}{\partial \phi}  \dot f(\phi) \right) \left[ u^*(k_1,\tau') u_R^*(k_2,\tau') u_R^*(k_3,\tau')\right] .
\end{align}
We can compute analytically these integrals with the same approximations we have already discussed above: in particular the Hubble parameter $H$ and the function $f(\phi)$ and its derivatives can be considered constant 
as a first approximation thanks to the slow-roll dynamics. The second approximation is about the cosmological evolution of the scale factor $a$. At leading order in slow-roll we have $a \simeq - 1/(H \tau)$. Finally we take the value of the functions $u$ and $u_s$ setting the slow-roll parameters and the ratio $H/M_{CS}$ equal to zero. This approximation is justified by the fact that these parameters are very small during inflation. Thus, as a result of the last approximation, we can use the mode function of a scalar field in a de Sitter space \eqref{u_integrable}
\begin{align}
u_s(k, \tau) = &\frac{i H}{M_{Pl} \sqrt{k^3}} (1+ik \tau) e^{-i k \tau} \mbox{ },\\
 u(k, \tau) = & \frac{i H}{\sqrt{2 k^3}} (1+ik \tau) e^{-i k \tau} \mbox{ },
\end{align}
where the different normalizations of the variables $\gamma$ and $\delta \phi$ come out from the study of the action of standard slow-roll inflationary models (see e.g. \cite{Maldacena:2003}).

With the prescriptions just explained we start now the explicit computation of the first integral $I_1$. It reads
\begin{equation}
I_1 = - 4 M^2_{Pl}\left( \prod_{i=1, 2, 3} \frac{H^2}{M^2_{Pl} 2 k^3} \right)\left(\frac{\partial}{\partial \phi}  f(\phi) \right) k^2_1 k^2_2 k^2_3\int_{- \infty}^0 d\tau' \tau'^3 e^{-i K_T \tau'} \mbox{ }, 
\end{equation}
where $K_T= k_1+k_2+k_3$. 

\noindent This integral can be performed by parts, after passing to the complex plane and performing a Wick rotation of the integration contour. We obtain
\begin{equation}
I_1 = - 4 M^2_{Pl} \left( \prod_{i=1, 2, 3} \frac{H^2}{M^2_{Pl} 2 k^3} \right)\left(\frac{\partial}{\partial \phi}  f(\phi) \right) k^2_1 k^2_2 k^2_3 \left( - \frac{3!}{K_T^4}\right)\mbox{ }. 
\end{equation}
We see that the final result is real. Then it does not give any contributions to the correlator \eqref{correlator_stt3}. For the same reason also the integrals $I_2$ and $ I_3$ do not give any contribution. The only integral which is not trivial is $ I_4$. Let us see its computation in details
\begin{equation}\label{I4}
 I_4 =  4 \frac{M^2_{Pl}}{H} \left( \prod_{i=1, 2, 3} \frac{H^2}{M^2_{Pl} 2 k^3} \right)\left(\dot \phi \frac{\partial^2}{\partial^2 \phi}  f(\phi) \right) \int_{- \infty}^0 \frac{d\tau'}{\tau'} (1+ik_1 \tau') (1+ik_2 \tau') (1+ik_3 \tau')e^{-i K_T \tau'} \mbox{ }.
\end{equation}
We write down the integral which appears in \eqref{I4} as
\begin{equation}
\int_{- \infty}^0 \frac{d\tau'}{\tau'} (1+ik_1 \tau') (1+ik_2 \tau') (1+ik_3 \tau')e^{-i K_T \tau'} \, ,
\end{equation}
which can be splitted into the sum of four integrals
\begin{equation}\begin{split}
&\left(\int_{- \infty}^0 \frac{d\tau'}{\tau'} e^{-i K_T \tau'} \right)+ \left(i K_T \int_{- \infty}^0 {d\tau'} e^{-i K_T \tau'}\right)- \left(\prod_{i \neq j} k_i k_j \int_{- \infty}^0 {d\tau'} \tau' e^{-i K_T \tau'}\right) +\\
&- \left(i k_1 k_2 k_3 \int_{- \infty}^0 {d\tau'} \tau'^2 e^{-i K_T \tau'}\right) \mbox{ }.
\end{split}\end{equation}
All the integrals, apart the first one, can be computed by parts and give a real contribution. For this reason they do not give any contribution to the correlator \eqref{correlator_stt3}. Instead, the first integral can be written in terms of the exponential integral $Ei(z)$ by promoting the real variable $\tau'$ to a complex variable and performing a Wick rotation of the integration contour (i.e. a change of variable $\tau'= -i \tau''$). It becomes:
\begin{equation}\label{red_integral}
\lim_{\tau \longrightarrow 0-} \int_{- i \infty}^{i \tau} \frac{d\tau''}{\tau''} e^{- K_T \tau''} \mbox{ }.
\end{equation}
The \textit{complex exponential integral} is defined as \cite{NIST:DLMF}:
\begin{equation}
Ei(z)= \int_{\infty}^{z} \frac{dz'}{z'} e^{- z'} \quad |Arg(z)|< \pi \mbox{ }.
\end{equation}
It is well defined for all complex numbers $z$ that are off the real negative axis. A good characteristic of this integral is that it is independent by the integration contour but it depends only by $z$. In particular it can be expressed in terms of the following series representation \cite{NIST:DLMF}:
\begin{equation}\label{exponential_integral}
Ei(z)= - \gamma - \ln{z} - \sum_{k=1}^\infty \frac{(-z)^k}{k \mbox{ }k!} \mbox{ }, 
\end{equation}
where $\gamma$ is the Euler-Mascheroni constant and $\ln{z}$ is the principal complex logarithm of the complex number $z$. This series converges for all $z$ that are not in the real axis. Applying formula \eqref{exponential_integral}, the integral \eqref{red_integral} becomes
\begin{equation}\label{red_integral2}
\lim_{ K_T \tau \longrightarrow 0-} \left[ - \gamma - \ln{(i K_T \tau)} - \sum_{k=1}^\infty \frac{(-i K_T \tau)^k}{k k!} \right]= - \gamma + \left(\lim_{K_T \tau \longrightarrow 0} \mbox{ }\ln{|K_T \tau|} \right)+ i \frac{\pi}{2}\mbox{ },
\end{equation}
where the $\ln|K_T \tau|$ in this case represents a real logarithm. Thus, at the end, we have:
\begin{equation}
\operatorname{Im}(I_4)=  4 \frac{M^2_{Pl}}{H} \left( \prod_{i=1, 2, 3} \frac{H^2}{M^2_{Pl} 2 k_i^3} \right)\left(\dot \phi \frac{\partial^2}{\partial^2 \phi} f(\phi) \right) \times \left(i \frac{\pi}{2}\right) \mbox{ }.
\end{equation}
If we substitute this result into Eq. \eqref{correlator_stt3} and we consider also the contributions of the permutations, we find the final result
\begin{equation}\begin{split} \label{correlator_stt4}
\langle  \gamma_{R}( \vec k_1) \gamma_{R}(\vec k_2)  \delta \phi(\vec k_3) \rangle = & (2 \pi)^3 \delta^3(\vec k_1+\vec k_2+\vec k_3)  4 \pi \frac{M^2_{Pl}}{H} \left( \prod_{i=1, 2, 3} \frac{H^2}{M^2_{Pl} 2 k_i^3} \right)\left(\dot \phi \frac{\partial^2}{\partial^2 \phi}  f(\phi) \right) \times\\
& \times (k_1+ k_2) ( \vec k_1 \cdot \vec k_2)  {\epsilon^R_{ij}}(\vec k_1) {\epsilon_R^{ij}}(\vec k_2) \mbox{ }.
\end{split}\end{equation}
Following the same steps, we are able  to compute the correlator $\langle \gamma_{L}( \vec k_1)  \gamma_{L}(\vec k_2)  \delta \phi(\vec k_3) \rangle$ as well. It is sufficient to substitute in the previous steps $R$ with $L$ and take a relative factor $-1$ due to the $\lambda_L= -\lambda_R$ relation in the interaction Lagrangian \eqref{L_stt3}. Thus, we have:
\begin{equation}\begin{split} \label{correlator_stt5}
\langle \gamma_{L}( \vec k_1) \gamma_{L}(\vec k_2)  \delta \phi(\vec k_3) \rangle =& - (2 \pi)^3 \delta^3(\vec k_1+\vec k_2+\vec k_3)  4 \pi \frac{M^2_{Pl}}{H} \left( \prod_{i=1, 2, 3} \frac{H^2}{M^2_{Pl} 2 k_i^3} \right)\left(\dot \phi\frac{\partial^2}{\partial^2 \phi}  f(\phi) \right) \times\\
& \times (k_1+ k_2) ( \vec k_1 \cdot \vec k_2)  {\epsilon^L_{ij}}(\vec k_1) {\epsilon_L^{ij}}(\vec k_2) \mbox{ }.
\end{split}\end{equation}

\noindent We can try to express the final result in a way in which we write explicitly the dependence over the wave numbers $k_i$. Because of momentum conservation, $\delta^{(3)} \left(\vec k_1+\vec k_2+\vec k_3 \right)=0$, the three momenta form a triangle in momentum space. For invariance under rotations we can put this triangle in the $(x,y)$-plane without losing any generality. It follows that a triangle can be constructed by
\begin{equation}
\vec k_1=k_1(1, 0, 0) \mbox{ ,} \quad \vec k_2=k_2 (\cos \theta, \sin \theta, 0), \mbox{ } \quad \vec k_3= k_3(\cos \Phi, \sin \Phi, 0) \mbox{ } ,
\end{equation}
where $\theta$ and $\Phi$ are the angles that the momenta $\vec k_2$ and $\vec k_3$ form respectively with the momentum $\vec k_1$. With these choices of the momenta, we can write down the explicit expressions of $L$ and $R$ polarization tensors. They read \cite{Soda:2011}
\begin{equation}
\epsilon_{ij}^{s}(\vec k_1)= \frac{1}{\sqrt 2}\begin{pmatrix} 0& 0 & 0 \\ 0 & 1 & i\lambda_s\\  0 & i \lambda_s & -1 \end{pmatrix} \mbox{ } ,
\end{equation}

\begin{equation}
\epsilon_{ij}^{s}(\vec k_2)= \frac{1}{\sqrt 2}\begin{pmatrix} \sin^2 \theta& - \sin \theta \cos \theta & - i \lambda_s \sin \theta \\ - \sin \theta \cos \theta & \cos^2 \theta & i \lambda_s \cos \theta\\ - i \lambda_s \sin \theta & i \lambda_s \cos \theta & -1 \end{pmatrix} \mbox{ } .
\end{equation}
 Thus through an explicit calculation we find:
\begin{equation}
\vec k_1 \cdot \vec k_2= k_1 k_2 \cos \theta , \qquad \epsilon_{ij}^{s}(\vec k_1) \epsilon^{ij}_{s}(\vec k_2)= \frac{1}{2} (1- \cos\theta)^2 \mbox{ },
\end{equation}
where $\theta$ is the angle between the two momenta $\vec k_1$ and $\vec k_2$. By the cosine theorem we can express this angle as a function of the three wave numbers $k_i$
\begin{equation}\label{angle}
\cos \theta= \frac{k_3^2- k_2^2 - k_1^2}{2 k_1 k_2} \mbox{ }.
\end{equation}
In the end the correlator \eqref{correlator_stt4} becomes
\begin{equation}\begin{split} \label{correlator_stt6}
\langle  \gamma_{R}( \vec k_1) \gamma_{R}(\vec k_2) \delta \phi(\vec k_3) \rangle =& (2 \pi)^3 \delta^3(\vec k_1+\vec k_2+\vec k_3) 4 \pi \frac{\dot \phi}{H} \left( \prod_{i=1, 2, 3} \frac{H^2}{M^2_{Pl} 2 k_i^3} \right)\left(M^2_{Pl}\frac{\partial^2}{\partial^2 \phi} f(\phi) \right)  \times\\
& \times (k_1+ k_2) k_1 k_2 \frac{\cos \theta (1-\cos \theta)^2}{2} \mbox{ }.
\end{split}\end{equation}

We can rewrite this correlator in a more convenient way in which the product of two power spectrum of PGW \eqref{power_spectrum_tensor} appears. It is sufficient to multiply and divide for $\sum_i k_i^3$ to obtain the final result
\begin{equation}\begin{split} \label{correlator_stt7}
\langle  \gamma_{R}( \vec k_1) \gamma_{R}(\vec k_2) \delta \phi(\vec k_3) \rangle =& (2 \pi)^3 \delta^3(\vec k_1+\vec k_2+\vec k_3) \frac{\pi}{64} \frac{\dot \phi}{H} \left( \sum_{i\neq j} \Delta_T(k_i) \Delta_T(k_j) \right)\left(H^2\frac{\partial^2}{\partial^2 \phi} f(\phi) \right)  \times\\
& \times \frac{(k_1+ k_2) k_1 k_2}{\sum_i k_i^3} \cos \theta (1-\cos \theta)^2 \mbox{ }\, .
\end{split}\end{equation}

\subsection{Main results} \label{5.4}

As we have understood in the previous subsections, the two graviton-one scalar bispectrum \eqref{correlator_stt7} is the most relevant for parity breaking signatures. Before starting to analyze such signatures we switch to the gauge invariant variable $\zeta$, using the local relation \eqref{zeta_non_linear}. Thus, in the coordinate space we have
\begin{equation} \label{local_red}
\langle \gamma_{R}( \vec x_1) \gamma_{R}(\vec x_2)  \zeta(\vec x_3) \rangle  \simeq -  \frac{H}{\dot \phi} \langle \gamma_{R}( \vec x_1)  \gamma_{R}(\vec x_2)  \delta \phi(\vec x_3) \rangle  \mbox{ ,}
\end{equation}
where we have not considered the contribution of the field redefinition coming from the non-linear part of the relation between $\zeta$ and $\delta \phi$. In fact this contribution represents a disconnected term. The symbol $\simeq$ means that we are evaluating the correlator at first order in the slow-roll parameters.

\noindent Thus passing in Fourier space and substituting Eq. \eqref{correlator_stt7} into Eq. \eqref{local_red} we have
\begin{equation}\begin{split} \label{correlator_stt8}
\langle  \gamma_{R}( \vec k_1) \gamma_{R}(\vec k_2)  \zeta(\vec k_3) \rangle  \simeq & - \frac{H}{\dot \phi} \langle  \gamma_{R}( \vec k_1)  \gamma_{R}(\vec k_2) \delta \phi(\vec k_3) \rangle= \\
&= - (2 \pi)^3 \delta^3(\vec k_1+\vec k_2+\vec k_3)  \frac{\pi}{64} \left(  \sum_{i\neq j} \Delta_T(k_i) \Delta_T(k_j) \right)\left(H^2 \frac{\partial^2 f(\phi) }{\partial^2 \phi}  \right) \times\\
& \times \frac{(k_1+ k_2) k_1 k_2}{\sum_i k_i^3} \cos \theta (1-\cos \theta)^2 \, ,
\end{split}\end{equation}
where we recall that $\theta$ is the angle between the two momenta $\vec k_1$ and $\vec k_2$.
 Proceeding with the same reasoning for computing the vertex $\langle \gamma_{L}( \vec k_1)  \gamma_{L}(\vec k_2)  \zeta(\vec k_3) \rangle$, we find:
\begin{equation} \label{correlator_stt9}
\langle \gamma_{L}( \vec k_1)  \gamma_{L}(\vec k_2)  \zeta(\vec k_3) \rangle= - \langle \gamma_{R}( \vec k_1) \gamma_{R}(\vec k_2) \zeta(\vec k_3) \rangle\, .
\end{equation}
Notice that we have obtained a result which differs for a sign in the passage from left to right gravitons. This result is not inconsistent and can be explained with a symmetry argument: if parity was a symmetry of the theory, there would be not difference between the statistics of L and R gravitons. And thus the correlators $\langle \gamma_R \gamma_R \zeta \rangle$ and $\langle \gamma_L \gamma_L \zeta \rangle$ would be equal. But the Chern-Simons term breaks parity symmetry and, as a consequence, we have a difference between the two bispectra. This is achievable e.g. by a sign difference, as it happens specifically in our case.~\footnote{Notice that this is what happens similarly in the analysis of~\cite{Shiraishi:2011}.}
\begin{figure}[h]
\centering
\includegraphics[scale=0.9]{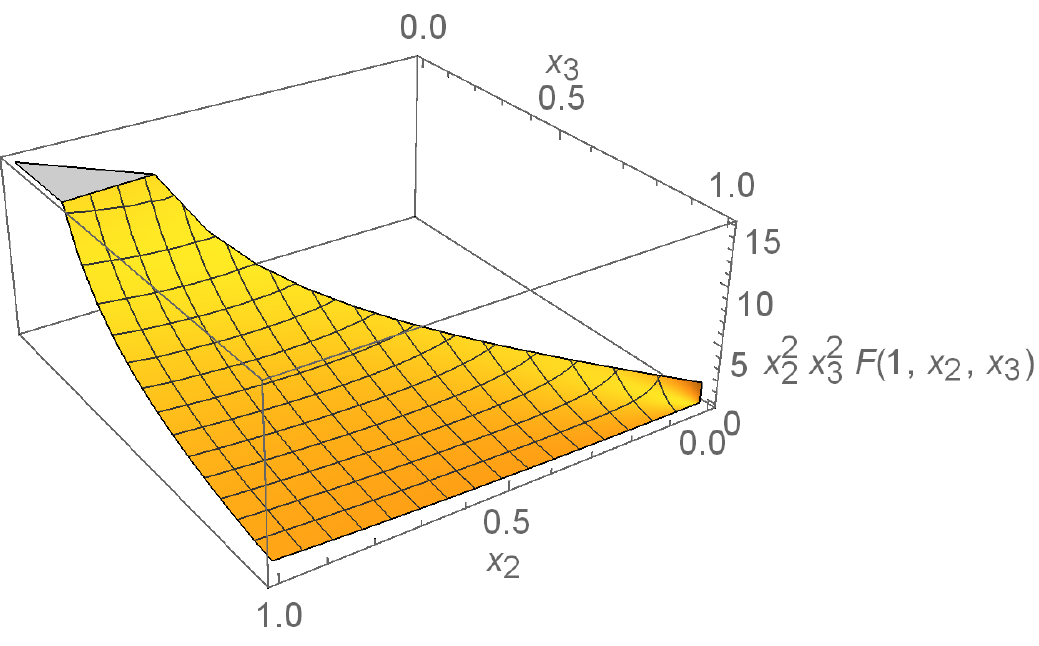} 
\caption{Plot of the shape of the correlator $\langle  \gamma_{R}( \vec k_1) \gamma_{R}(\vec k_2)  \zeta(\vec k_3) \rangle$. The quantity $F(1,x_2,x_3) x^2_2 x_3^2$ in terms of $x_2 = \frac{k_2}{k_1}$ and $x_3 = \frac{k_3}{k_1}$ is shown. The figure is normalized to have value $1$ for equilateral configurations $x_2=x_3=1$.}\label{plot_shape}
\end{figure}
If we take the dependence of the bispectrum  \eqref{correlator_stt8} over the momenta $k_i$'s, we obtain the shape function
\begin{equation}\label{shape}
F(k_1, k_2, k_3) =  \left(\sum_{i \neq j} \frac{1}{k^3_i k_j^3} \right) \frac{(k_1+ k_2) k_1 k_2}{\sum_i k_i^3} \cos \theta (1-\cos \theta)^2,
\end{equation}
where $\cos \theta$ is defined in Eq.~\eqref{angle}. As it is customary for scale-independent bispectra, the shape function is plotted as the quantity $F(1, x_1, x_2) x_2^2 x_3^2$ in terms of the variables $x_2=k_2/k_1$ and $x_3=k_3/k_1$. From Fig. \ref{plot_shape} we see that the shape function peaks when $x_2= 1$ and $x_3=0$. This corresponds to the so-called squeezed limit, in which the momenta of the gravitational waves $k_1$, $k_2$ are much larger than the momentum $k_3$ of the scalar perturbation. 

At this point we can start to analyze the parity breaking signatures induced by our result \eqref{correlator_stt8} in the bispectra we are analysing. For this purpose, we can define a parity breaking coefficient\footnote{This coefficient is not properly an amplitude of primordial non-Gaussianity $f_{NL}$, but an estimate of how large parity breaking is in our bispectra. In any case one can define separately for L and R polarizations a coefficient of non-Gaussianity $f^{R,L}_{NL}$ as
\begin{equation}
\label{amplitudes}
f^{R,L}_{NL} = \frac{\langle \gamma_{R,L}( k)  \gamma_{R,L}(k)  \zeta(k) \rangle}{\Delta_T^2(k)}\, ,
\end{equation}
finding 
\begin{equation}
\label{amplitudese}
f^{R}_{NL} =\frac{\pi}{256}H^2 \frac{\partial^2 f(\phi) }{\partial^2 \phi},
\end{equation}
and the same result for $f^{L}_{NL}$ apart a sign difference. From this definition it is clear that we have normalized the L and R bispectra using the tensor power-spectrum \eqref{power_spectrum_tensor}.
An alternative definition can be adopted by normalizing with the scalar power-spectrum, which would simply give the same result as above rescaled by the inverse of the tensor-to-scalar perturbation ratio. In this context we do prefer the definition~(\ref{amplitudes}) because the parity breaking signatures arise in the tensor sector. 
Using the result~(\ref{amplitudes}) and Eq. \eqref{Pi} we can link such coefficient of non-Gaussianity to the coefficient $\Pi$ finding
\begin{equation}
\label{fNLRPi}
f^{R}_{NL} = \frac{25}{24576} \Pi \simeq 10^{-3} \Pi\mbox{ .}   
\end{equation}
}
\begin{equation}\label{pi}
\Pi = \frac{\langle \gamma_{R}(\vec k)  \gamma_{R}(\vec k)  \zeta(\vec k) \rangle_{TOT} - \langle \gamma_{L}(-\vec k)  \gamma_{L}(-\vec k)  \zeta(-\vec k) \rangle_{TOT}}{\langle \gamma_{R}(\vec k)  \gamma_{R}(\vec k)  \zeta(\vec k) \rangle_{TOT} + \langle \gamma_{L}(-\vec k)  \gamma_{L}(-\vec k)  \zeta(-\vec k) \rangle_{TOT}},
\end{equation}
where the suffix ``$TOT$" stands for the total bispectra, when we include the contribution of standard gravity (see Ref. \cite{Maldacena:2003}). In our conventions the latter contribution reads like 
\begin{equation}\begin{split}\label{stand_contrib}
\langle \gamma_{s}( \vec k_1)  \gamma_{s'}(\vec k_2)  \zeta(\vec k_3) \rangle'_{stand} = &\delta_{ss'} \left(  \sum_{i\neq j} \Delta_T(k_i) \Delta_T(k_j) \right)  \times\\
& \times \frac{(1-\cos\theta)^2}{256 \sum_i k_i^3} \left(-\frac{1}{4} k_3^3 + \frac{1}{2} k_3 (k_1^2 + k_2^2)+ \frac{4 k_1^2 k_2^2}{k_1+k_2+k_3} \right),
\end{split}\end{equation}
where the prime means that a Dirac delta is understood. 

In Eq.~\eqref{pi} the numerator represents the difference between R and L contributions to the bispectrum $\langle \gamma \gamma \zeta \rangle$ which ultimately will determine the amplitude of observable quantities, such as, e.g., CMB bispectra involving (mixed) temperature fluctuations (T) and polarization (E and B) correlators, like, e.g., 3-point correlations of the type $\langle T BB\rangle$ (see discussion in the conclusions).
In defining the amplitude $\Pi$ we have evaluated the bispectra in the equilateral configuration of the three momenta. This is a standard convention which is adopted in literature (see e.g. \cite{Planck_primordial_NG_2016}). To evaluate the quantity $\Pi$ notice that in the model considered\footnote{In Eq. \eqref{correlator_stt9} we have showed that $\langle \gamma_R (\vec k_1) \gamma_R(\vec k_2)\zeta(\vec k_3) \rangle=- \langle  \gamma_L(\vec k_1) \gamma_L(\vec k_2) \zeta(\vec k_3)\rangle$. Eq. \eqref{vec=-vec} follows after noticing that the bispectrum \eqref{correlator_stt8} depends only on the modulus of the momenta $\vec k_i$'s and not by their direction (In fact $\cos{\theta}$ can be expressed in terms of the $k_i$'s as showed in Eq. \eqref{angle}).}
\begin{equation}\label{vec=-vec}
\langle \gamma_R (\vec k_1) \gamma_R(\vec k_2)\zeta(\vec k_3) \rangle=- \langle  \gamma_L(-\vec k_1) 
\gamma_L(-\vec k_2) \zeta(-\vec k_3)\rangle \, .
\end{equation}
Thus, using Eq.~\eqref{stand_contrib} and our results, Eq.~(\ref{correlator_stt8}) and (\ref{correlator_stt9}), we find
\begin{equation}\label{Pi}
\Pi = \frac{96 \pi}{25}H^2 \frac{\partial^2 f(\phi) }{\partial^2 \phi} \mbox{ .}
\end{equation}
Now, let us comment about the result we have obtained in Eq. \eqref{Pi}. The parity-breaking amplitude $\Pi$ depends on the strength of the second derivative of the coupling $f(\phi)$ w.r.t. the inflaton which is something new with respect to the power-spectrum case.~\footnote{Arrived at this point one might ask whether the appearance of the second derivative is something of inevitable. In fact it is interesting to notice that if the first derivative of the coupling $f(\phi)$ w.r.t. the inflaton would be {\it exactly} a constant in the first term of Eq.~(\ref{L_stt}), then one would obtain a vanishing result for the correlators of interest. This confirms the result that we find in Eq.~(\ref{correlator_stt8}), namely that these correlators are actually sensitive to the second derivative $f(\phi)$ w.r.t. the inflaton field.}
In fact in the latter case only the first derivative of the coupling $f(\phi)$ appeared (see Eq.~(\ref{theta}) and~(\ref{relation_partial})). Thus, apparently we can say that our result is independent of the parity breaking in the power-spectrum case which is labelled by the parameter $\Theta$ in Eq. \eqref{theta}. However, a theoretical constraint occurs as a consequence of some of assumptions we have adopted in our computations: in fact, we have assumed that the coupling $f(\phi)$ and all its derivatives do not change significantly in time due to slow-roll dynamics. This has allowed us to consider the Chern-Simons mass as a constant parameter and to put derivatives of $f(\phi)$ out of the integrals in using the in-in formalism formula, Eq. \eqref{master}.  In particular, let us define a dimensionless parameter, $\xi$, which quantifies the time dependence of Chern-Simons mass
\begin{equation}
\xi = \frac{\dot M_{CS}}{H M_{CS}} \mbox{ .}
\end{equation}
If $\xi \ll 1$, the rate of variation of the Chern-Simons mass is much smaller then the Hubble parameter. 
By a direct computation, using the definition of the Chern-Simons mass, Eq.~\eqref{mass_CS}, we find
\begin{equation}\label{value_xi}
\xi = \epsilon - \eta + \sqrt{2 \epsilon} M_{Pl} \frac{f''(\phi)}{f'(\phi)} \mbox{ ,}
\end{equation}
where the prime $'$ denotes derivative w.r.t. the inflaton, $\epsilon$ is defined in Eq. \eqref{epsilon} and $\eta$ is the other slow-roll parameter
\begin{equation}
\eta = M^2_{Pl} \frac{V''}{V} \simeq -\frac{\ddot\phi}{H \dot\phi}+ \frac{1}{2} \frac{\dot{\phi^2}}{H^2} M^{-2}_{Pl}  \label{eta} \mbox{ .}
\end{equation}
In slow-roll models of inflation, we know that slow-roll parameters must be much smaller than 1. Thus, from Eq. \eqref{value_xi}, we need only to require
\begin{equation}\label{constraint}
\sqrt{2 \epsilon} M_{Pl} \frac{f''(\phi)}{f'(\phi)} \ll 1 
\end{equation}
in order to neglect the time dependence of the Chern-Simons mass (apart from fine-tuned cancellations).  In addition we can link $f'(\phi)$ to the parameter $\Theta$ using Eqs.~\eqref{theta} and \eqref{relation_partial}. We have
\begin{equation}\label{f_to_theta}
M_{Pl} f'(\phi)= M_{Pl} \frac{\dot f}{\dot \phi} =\frac{1}{4 \sqrt{2} \pi} \frac{\Theta}{\sqrt{\epsilon}} \frac{M^2_{Pl}}{H^2} \mbox{ }.
\end{equation}
If we insert Eq. \eqref{f_to_theta} into Eq. \eqref{constraint}, we obtain the following theoretical constraint 
\begin{equation}\label{constraint_2}
H^2 f''(\phi) \ll \frac{1}{8 \pi} \frac{\Theta}{\epsilon} \mbox{ .}
\end{equation}
From the last equation and Eq.~\eqref{Pi} we understand that in a model in which the ratio $\Theta/ \epsilon$ is sufficiently large (e.g. greater than $10^2$), relatively high values of $\Pi$ are compatible with the theory. This can happen also in the case in which parity breaking in the power spectra of the tensor perturbations is very small (i.e. $\Theta = 10^{-1}$ or smaller) and not measurable. This result can be interesting, because it shows that a significant parity breaking signature can arise, at least, in the bispectrum $\langle \gamma \gamma \zeta \rangle$, even if parity breaking in the power spectrum is very small.

It is also interesting to notice that such a result arises because in this model the correlators $\langle  \gamma \gamma \zeta \rangle$ are parametrically enhanced w.r.t. to the standard single-field models of slow-roll inflation. Looking at Eq.~(\ref{correlator_stt8}) and Eq.~(\ref{stand_contrib}) we see that, in terms of the strength of the bispectra (or equivalently in terms of the amplitudes $f^{R,L}_{NL}$)
\begin{equation}\label{chern_vs_stand}
\langle  \gamma_R \gamma_R \zeta \rangle \simeq 0.1\ H^2 f''(\phi)\, \langle \gamma_{R} \gamma_{R} \zeta \rangle_{stand}\, ,
\end{equation}
where $H^2 f''(\phi)$ can be at most of the order of $\Theta/\epsilon$. It is then enough to consider Eq.~(\ref{Pi}) and recall that in this model $\langle  \gamma_R \gamma_R \zeta \rangle=- \langle  \gamma_L \gamma_L \zeta \rangle$.

\section{Comments about the squeezed limit}\label{6}

First of all let us notice that a useful cross-check of our result~(\ref{correlator_stt8}), and~(\ref{correlator_stt9}), comes from considering its squeezed limit, in which we take the momentum of the scalar fluctuation $\vec k_3 = \vec k_L\simeq 0$ and the other two momenta of the tensor fields such that $\vec k_1 = \vec k_S \simeq -\vec k_2$. In this limit we know that the bispectrum can be computed by considering the effect of the long-wavelength mode on the short-wavelength ones. Briefly speaking, when the scalar fluctuation has a wavelength much larger than those of the two tensor modes it exits the horizon much before than the two tensors do. Then in this case the scalar perturbation behaves like a background that modulates the tensor power-spectrum. Thus in formula our bispectrum becomes (see e.g. \citep{Maldacena:2003})
\begin{equation}\label{modulation}
\langle  \gamma_{R}( \vec k_S) \gamma_{R}(-\vec k_S)  \zeta(\vec k_L) \rangle|_{squeezed} = - \Delta_S(k_L) \frac{1}{H} \frac{d}{dt} \Delta_T^R(k_S)\, ,
\end{equation}
where
\begin{equation}
\Delta_S(k) = \frac{H^2}{4 \epsilon M^2_{Pl} k^3}
\end{equation}
is the scalar power spectrum of slow-roll models of inflation and $\Delta_T^R(k)$ is the right-handed tensor power-spectrum as in Eq. \eqref{delta_expansion}. By a direct computation we find 
\begin{equation}\label{comput_consis}
\frac{1}{H}\frac{d}{dt} \Delta_T^R(k_S)\simeq \left[- 2 \epsilon + \frac{\pi}{4}\left(\frac{\dot M_{CS}}{M^2_{CS}}\right) \right] \times \frac{\Delta_T(k_S)}{2} \, .
\end{equation}
Here $\simeq$ refers to the dominant contributions in slow-roll parameters and in $(H/M_{CS})$, and $\Delta_T(k)$ is the tensor power spectrum as in Eq. \eqref{power_spectrum_tensor}. Since we are interested only in contributions of the parity breaking part of the bispectrum, we consider in Eq. \eqref{comput_consis} only the term that depends on the Chern-Simons mass. Using the definition \eqref{mass_CS}  we find
\begin{equation}
\frac{\dot M_{CS}}{M^2_{CS}} \simeq - 16 H^2 f''(\phi) \epsilon\, ,  
\end{equation}
where the prime $'$ denotes derivative w.r.t. the inflaton field. Here, similarly as above, we have neglected another term that is proportional to the ratio $H/M_{CS}$ which is much smaller than 1 in our theory. This term, in fact, has not been considered in our computation of the bispectrum performed in the previous section. 

Thus, gathering all together in Eq. \eqref{modulation}, the bispectrum \eqref{correlator_stt8} in the squeezed limit becomes
\begin{equation}\label{modulation2}
\langle  \gamma_{R}( \vec k_S) \gamma_{R}(-\vec k_S)  \zeta(\vec k_L) \rangle|_{squeezed} =  \frac{\pi}{8} H^2 f''(\phi) \Delta_T(k_L) \Delta_T(k_S) \mbox{ .}
\end{equation}
This result (which here has been obtained in the case of a parity-breaking theory) corresponds to the so called ``consistency conditions'' (see, e.g.~\cite{Maldacena:2003}). It is easy to verify that if we take our complete result~(\ref{correlator_stt8}) and expand it around the squeezed limit, then at leading order in $(k_L/k_S)$ we obtain exactly Eq.(\ref{modulation2}).
 
Secondly,  it  has been shown in general that the result one obtains in the strict squeezed limit corresponds to a gauge artifact (see Refs. \cite{Tanaka:2011aj,Dai:2013kra,Creminelli:2013bis,Creminelli:2013,Pajer:2013ana,Khoury:2014,Mirbabayi:2015,Bordin:2016,Cabass:2017}). 
It is possible to show that one can perform a coordinate redefinition, passing from co-moving coordinates to Conformal Fermi coordinates. The latter is the coordinate frame of an observer that ``sees'' inflation in the background perturbed by the long-wavelength scalar mode $\zeta(\vec k_L)$. After performing this coordinate transformation, our bispectrum would become (see Refs.~\cite{Creminelli:2011,Bordin:2016, Cabass:2017})
\begin{equation}\label{transf_squeez2}
\langle  \gamma_{R}( \vec k_S) \gamma_{R}(-\vec k_S)  \zeta(\vec k_L) \rangle|_{squeezed} =\mathcal{O}\left(\frac{k_L^2}{k_S^2}\right) \times\Delta_T(k_L) \Delta_S(k_S)  \mbox{ .}
\end{equation}
The quantity $\mathcal{O}\left( k^2_L/k^2_S\right)$\footnote{We have not computed explicitly the coefficient which multiplies the factor $k_L^2/k_S^2$, because it is not the purpose of this paper (for such a computation in the case of the scalar bispectrum, see~\cite{Cabass:2017}). Our aim, in fact, is to show that one has to pay attention when constraining with observations a bispectrum from single-clock inflation that peaks in the squeezed configuration.} is a correction term to the consistency relation that must be there at quadratic-order in $k_L/k_S$.

After this transformation, residual gauge modes are completely removed and so we remain with the correct squeezed amplitude of the bispectrum. We can say that \eqref{transf_squeez2} is the bispectrum that one expects to measure in the very squeezed limit. An analogous argument is valid (apart for a sign difference) also for the squeezed bispectrum $\langle  \gamma_{L}( \vec k_S) \gamma_{L}(- \vec k_S)  \zeta(\vec k_L) \rangle$.

A problematic consequence of this fact appears when one tries to compute the signal-to-noise ratio for the bispectrum $B(k_1, k_2, k_3)$. This in formula reads 
\begin{equation}\label{SN}
\frac{S}{N} = \left(\sum_{l_1 \leq l_2\leq l_3} \frac{B^2_{l_1,l_2,l_3}}{C_{l_1} C_{l_2} C_{l_3}}\right)^{\frac{1}{2}} \approx \left(\int dx_2 dx_3 B^2(1, x_2, x_3) x_2^4 x_3^4 \right)^{\frac{1}{2}} \mbox{ .}
\end{equation}
Here the integration over $x_3$ goes from $x_{min}= l_{min}/l_{max}$ to $1$, while the integration over $x_2$ goes from $1-x_3$ to $1$. Since for {\it Planck} we can take  $l_{min}={2}$ and $l_{max} \approx 2000$ (see Ref. \cite{Planck_primordial_NG_2016}), $x_{min}= 10^{-3}$. In the squeezed limit our physical bispectrum goes as in Eq. \eqref{transf_squeez2}. But this is valid only when $x_3$ is sufficiently small. For example, when $x_3$ is greater than $10^{-1}$ we are already too far from the squeezed configuration to use the expression \eqref{transf_squeez2}. On the contrary, in this case we have to use the result \eqref{correlator_stt8}. For this reason it is interesting to compare the signal-to-noise ratio calculated in two different cases: in a first case we substitute the bispectrum \eqref{correlator_stt8} into Eq. \eqref{SN} without taking into consideration spurious signatures that come from the squeezed configurations; in a second case, instead, we split the integration over $x_3$ into two parts: when  $x_3<10^{-1}$ we use the squeezed expression \eqref{transf_squeez2} for the bispectrum\footnote{We have verified that our conclusions are not sensitive to the precise value of the term ${\mathcal O}(k_L^2/k_S^2)$ since in the squeezed limit ($x_3\approx 0$) the integrand of Eq. \eqref{SN} is suppressed as $x_3^2$ and thus it gives a negligible contribution over squeezed configurations.}, instead when $x_3 \geq 10^{-1}$ we use expression \eqref{correlator_stt8}. If we compare the two $S/N$ ratios we obtain that in the second case we lose approximately $50 \%$ of the signal that we would have if we take into considerations also spurious signatures. This implies that to produce the same $S/N$ ratio, the primordial parity breaking amplitude $\Pi$ should increase by a factor two w.r.t to the case where we do consider spurious effects in the squeezed limit, something which however is not difficult to obtain within the parameter space of the model. Of course, this is just an estimate, because one should define better the value of $x_3$ at which the very squeezed limit ends. 


\section{Conclusions}\label{7}
In this paper we have studied parity breaking signatures in the primordial bispectra induced by a Chern-Simons gravitational term coupled to the inflaton field. This term introduces parity breaking in the theory, polarizing  PGW into Left and Right chiral eigenstates and leaving scalar modes unchanged. The starting point of the theory has been the requirement of the absence of ghost-fields. Our analysis was motivated by the fact that in such a case parity breaking signatures in power spectra have to be suppressed. We have seen that an asymmetry among the bispectra $\langle \gamma \gamma \zeta \rangle$ for each polarization state arises as the only  breaking signature that is not suppressed as the power spectra case. Such a signature receives a relevant contribution from the squeezed configuration (the momenta of the two tensor perturbations being much greater than the one of the scalar perturbation). In addition the strength of the signature is partially constrained by theoretical assumptions of the theory itself. However, this constraint tends to be alleviated when the slow-roll parameter $\epsilon$ is sufficiently smaller than $\Theta$, the coefficient which labels parity breaking in the power spectra (see Eq.~(\ref{constraint_2})). In this case we can have a relatively large parity breaking in the $\langle \gamma \gamma \zeta \rangle$ bispectra even if parity breaking in the tensor power spectrum is very low. Thus, briefly speaking, we have presented a model in which a possible parity breaking signature can be reached in the bispectrum statistics of primordial perturbations. We paid attention to the fact that the parity breaking signature is proportional to $f''(\phi)$. {For this reason only a non-minimal coupling function (i.e. $f(\phi)$ which is not just proportional to $\phi$) is able to produce such a signature. Then, if experimental observations will indicate such a signature, we would have an evidence that a non-minimal coupling between the Chern-Simons term and the inflaton field should be taken into consideration.}

Of course a dedicated analysis should be performed regarding which observables can be used to constrain such a signature and their precise sensitivity to the parity breaking amplitude $\Pi$, Eq.~(\ref{Pi}), for the bispectra considered in this paper~\cite{BOS}. A possibility would be to consider (mixed) CMB bispectra among temperature T and E/B polarization modes. For example, let us write in a schematic way the harmonic coefficients of the CMB temperature and polarization fields as 
\begin{eqnarray}
a_{\ell m}^{X}&\propto& \int d^3k\, Y^*_{\ell m}(\hat k) \, T^\zeta_{X}(k) \,\left[\zeta(\vec k)+(-1)^{\ell} \zeta(-\vec k) \right]  \,, \nonumber \\
a_{\ell m}^{B}&\propto&\int d^3k\, _{-2}Y^*_{\ell m}(\hat k)\, T^{\gamma}_{B}(k)\,\left[ \gamma_R(\vec k)-(-1)^{\ell} \gamma_L(-\vec k) \right]  ,
\end{eqnarray}
where $X$ is either T or E, and we have taken into account that T anisotropies are sourced mainly by scalar perturbations, while B polarization modes are sourced only by primordial gravitational waves. Thus, e.g, we would obtain that the following bipectrum\footnote{Notice that, a priori, one could use also a correlator without $B$ modes, e.g. $\langle TTT \rangle$, to link our prediction to the observations. The possible advantage of using correlators including $B$-modes as $\langle BBT \rangle$ instead of $\langle TTT \rangle$ to constrain  $\langle \gamma \gamma \zeta \rangle$ follows by estimating the ratio between the two S/N ratios. In fact following the same reasoning of  Ref. \cite{Meerburg:2016} and considering only a cosmic variance limited experiment  we would have

\begin{equation}
\frac{(S/N)^2_{BBT}}{(S/N)^2_{TTT}} \sim \frac{1}{r^2}  \left( \frac{\langle \gamma \gamma \zeta \rangle}{\langle \gamma \gamma \zeta\rangle} \right)^2 \sim \frac{1}{r^2} \, .
\end{equation}
From this equation it follows that for enough low values of r, $\langle BBT \rangle$ would provide better constraints than $\langle TTT \rangle$ on $\langle \gamma \gamma \zeta\rangle$. See the discussion in~\cite{Meerburg:2016}.}
\begin{equation}
\langle   a_{\ell_1 m_1}^{B} a_{\ell_2 m_2}^{B} a_{\ell_3 m_3}^{T} \rangle \sim \left[ \langle \gamma_R(\vec k_1) \gamma_R(\vec k_2) \zeta(\vec k_3) \rangle+(-1)^{\ell_1+\ell_2+\ell_3} \langle \gamma_L(-\vec k_1) \gamma_L(-\vec k_2) \zeta(- \vec k_3) \rangle \right]
\end{equation}
would be proportional to the parity breaking amplitude $\Pi$ for the specific multipole location $\ell_1+\ell_2+\ell_3=$odd. Namely it turns out 
\begin{equation}
\label{estimate}
\langle   a_{\ell_1 m_1}^{B} a_{\ell_2 m_2}^{B} a_{\ell_3 m_3}^{T} \rangle \sim  \left[ \langle \gamma_R \gamma_R \zeta \rangle- \langle \gamma_L \gamma_L \zeta \rangle \right] \propto 2 \langle  \gamma_R \gamma_R \zeta \rangle \propto f^R_{NL}\, F(k_1,k_2,k_3) \propto \Pi \,F(k_1,k_2,k_3)\, , 
\end{equation}
where we have used Eq.~(\ref{vec=-vec}) and the fact that in the model considered mixed correlators, like $\langle  \gamma_L \gamma_R \zeta \rangle$ vanish. In Eq.~(\ref{estimate}) $F(k_1,k_2,k_3)$ is the shape of the bispectrum $\langle \gamma_R \gamma_R \zeta \rangle$, Eq.~(\ref{shape}), while $f^R_{NL}$ is its non-Gaussianity amplitude defined in Eq.(\ref{amplitudese}), and related to  $\Pi$ by Eq.~(\ref{fNLRPi}). 

To conclude, let us discuss the possibility to relax some of the conditions we have adopted so far to investigate this model. First of all we notice that if the Chern-Simons mass can significantly vary in time during inflation, the constraint \eqref{constraint_2} can be strongly relaxed, leaving $f''(\phi)$, and thus any parity breaking signature for the bispectrum $\langle\gamma \gamma \zeta \rangle$, theoretically unconstrained. However this requires a deep reformulation of the analysis of the power-spectrum case, because in such a scenario the equation of motion \eqref{eomCS_3} will change drastically. Another interesting possibility can be to replace the inflaton field $\phi$ in the coupling function with another dynamical field $\chi$ different from the inflaton, in such a way to deal with a $f(\chi)$ coupling. In this case this additional field is not forced to follow a slow-roll dynamics as the inflaton and then, again, the  constraint \eqref{constraint_2} no longer holds. Moreover, in this case,  the model would not be single-clock inflation and so the discussion about the consistency relations provided in Section \ref{6} are no more valid, in particular regarding a potential signal-to-noise ratio loss. 

\section*{Aknowledgments}
It is a pleasure to thank Maresuke Shiraishi for many and insightful discussions and for useful correspondence. We would also like to thank Dario Cannone for valuable discussions and for initial collaboration on this work. We also thank Lorenzo Bordin, Giovanni Cabass, Michele Liguori and Sabino Matarrese for valuable discussions. We thank Angelo Ricciardone, Gianmassimo Tasinato, Lorenzo Bordin and Giovanni Cabass for reading the draft and for useful comments. We acknowledge financial support by ASI Grant 2016-24-H.0.
N.B. acknowledges partial financial support by the ASI/INAF Agreement I/072/09/0 for the Planck LFI Activity of Phase E2.

\appendix

\section{Expression of the Effective Lagrangian in terms of the Weyl-tensor} 
\label{appendix_A}
\noindent Let us start from the Lagrangian 
\begin{equation}\label{effective_lagrangian_appendix}
\Delta \mathcal{L}= \sqrt{g} \left(f_1 R^2 + f_2 R_{\mu \nu} R^{\mu \nu} + f_3 R_{\mu \nu \rho \sigma} R^{\mu \nu \rho \sigma} \right) + f_4 \epsilon^{\mu \nu \rho \sigma} {R_{\mu \nu}}^{\kappa \lambda} R_{\rho \sigma \kappa \lambda} . 
\end{equation}
If we use the definition of the Weyl tensor \eqref{Weyl_tensor} and we do explicitly tensor contractions we can show that (see Ref. \cite{Grumiller:2007rv})
\begin{equation}\label{C_to_R}
\epsilon^{\mu \nu \rho \sigma} {C_{\mu \nu}}^{\kappa \lambda} C_{\rho \sigma \kappa \lambda} = \epsilon^{\mu \nu \rho \sigma} {R_{\mu \nu}}^{\kappa \lambda} R_{\rho \sigma \kappa \lambda} 
\end{equation}
and
\begin{equation}
C_{\mu \nu \rho \sigma} C^{\mu \nu \rho \sigma} = R_{\mu \nu \rho \sigma} R^{\mu \nu \rho \sigma} - 2 R_{\mu \nu} R^{\mu \nu} + \frac{1}{3} R^2 . 
\end{equation}
Thus, if we define 
\begin{align}
h_1= & f_1 - \frac{1}{3} f_3 , \\
h_2 = & f_2 + 2 f_3 ,\\
h_3= & f_3 ,
\end{align}
then Lagrangian \eqref{effective_lagrangian_appendix} can be rewritten as:
\begin{equation}
\Delta \mathcal{L}= \sqrt{g} \left(h_1 R^2 + h_2 R_{\mu \nu} R^{\mu \nu} + h_3 C_{\mu \nu \rho \sigma} C^{\mu \nu \rho \sigma} \right) + f_4 \epsilon^{\mu \nu \rho \sigma} {C_{\mu \nu}}^{\kappa \lambda} C_{\rho \sigma \kappa \lambda} .
\end{equation}
If we rename the $h_i$'s as $f_i$'s, then Eq. \eqref{effective_lagrangian_weyl} holds.


\begingroup 
  \makeatletter
  \let\ps@plain\ps@empty
  \makeatother
  \bibliography{Bibliography5}
  \bibliographystyle{JHEP} 
\endgroup

\end{document}